\documentclass[reprint,prb,floatfix,twocolumn,superscriptaddress,nofootinbib,longbibliography,aps]{revtex4-2}
\usepackage{amsmath,graphicx,amssymb,braket,enumitem,mathtools,dsfont}
\usepackage[colorlinks=true,citecolor=blue,linkcolor=red]{hyperref}

\usepackage{titlesec}
\usepackage{booktabs}
\usepackage{siunitx}
\usepackage{bbold}
\usepackage{relsize}
\usepackage{tikz}
\usetikzlibrary{matrix,shapes.callouts}
\usepackage{subfig}

\usepackage{caption}
\titlespacing{\subsection}
    {0pt}{9pt}{6pt}

\graphicspath{{Figs}}

\begin{document}

\title{Multi-level charge fluctuations in a Si/SiGe double quantum dot device}
\author{Dylan Albrecht}
\affiliation{Sandia National Laboratories, Albuquerque NM, USA}
\author{Feiyang Ye}
\affiliation{University of Rochester, Rochester NY, USA}
\author{N. Tobias Jacobson}
\affiliation{Sandia National Laboratories, Albuquerque NM, USA}
\author{John M. Nichol}
\affiliation{University of Rochester, Rochester NY, USA}

\date{June 2025}

\begin{abstract}
Discrete charge fluctuations, routinely observed in semiconductor quantum dot devices, may contribute significantly to device drift and errors resulting from qubit miscalibration. Understanding the nature and origins of these discrete charge fluctuations may provide insights into material improvements or means of mitigating charge noise in semiconductor quantum dot devices. In this work, we measure multi-level charge fluctuations present in a Si/SiGe double quantum dot device over a range of device operating voltages and temperatures. To characterize the parameter-dependent dynamics of the underlying fluctuating degrees of freedom, we perform a detailed analysis of the measured noise timeseries. 
We perform algorithmically assisted drift detection and change point detection to detrend the data and remove a slow fluctuator component, as a preprocessing step.  We perform model comparison on the post-processed time series between different $n$-level fluctuator ($n$LF) factorial hidden Markov models (FHMMs), finding that although at most sweep values the independent pair of 2LFs model would be preferred, in a particular region of voltage space the 4LF model outperforms the other models, indicating a conditional rate dependence between the two fluctuators. By tracking fluctuator transition rates, biases, and weights over a range of different device configurations, we estimate gate voltage and conductivity sensitivity. In particular, we fit a phenomenological, detailed balance model to the extracted independent 2LFs rate data, yielding lever arm estimates in the range of $-2 \mu$eV/mV up to $4 \mu$eV/mV between the two 2LFs and nearby gate electrodes. We expect that these characterization results may aid in subsequent spatial triangulation of the charge fluctuators.

\end{abstract}

\maketitle
\clearpage

\section{Introduction}
\noindent
Random telegraph noise (RTN) is a significant noise source in semiconductor quantum devices that contributes to a widely observed $1/f$ charge noise spectrum~\cite{Machlup1954, Freeman2016}.  This noise is commonly attributed to two-level charge flucutators (TLFs) jumping between localized trap sites, likely located at an interface~\cite{Grasser2012}, as in the oxide interface in Si/SiO$_2$ devices, and is present in leading candidate heterostructures~\cite{Freeman2016, Wang2025}. 
The majority of $1/f$ charge noise and the associated decoherence could be explained by a collection of TLFs~\cite{Shehata2023}, and possibly a very small number of strongly contributing TLFs~\cite{Mehmandoost2024}. Charge noise is commonly the primary limiter to device performance~\cite{Yoneda2018}. Understanding the nature of this noise source by characterizing, controlling, and triangulating, as well as using techniques to alleviate the effects of TLFs, is important for the future of this qubit technology (see~\cite{Burkard2023} for a recent review on spin qubits).  

There have been numerous efforts to characterize the extent of spatial correlations due to charge noise in qubit devices using Green's function methods~\cite{Cheng2025} and cross correlations~\cite{Rojas-Arias2023, Yoneda2023}, as well as wavelet analysis for spatio-temporal correlations~\cite{Seedhouse2025}.
A possible understanding of these correlations being the interaction via elastic strain~\cite{Mickelsen2023}, displaying a similar exponential drop off of correlation. Additionally, there have been efforts toward directly measuring the location, transition rates, and correlation lengths of TLFs in these types of systems~\cite{Graaf2015,Cowie2024}. Information to directly inform fabrication would be highly valuable in order to mitigate the creation of TLF noise and the downstream effects. Recent efforts to mitigate charge noise range from adjusting the device fabrication process~\cite{Wuetz2023} to optimizing device operation in the presence of noise~\cite{Choi2024}.

Analyzing and manipulating clearly observable TLF noise in the time domain provides detailed information about the device-TLF interaction and sensitivity. This type of analysis enables the determination of temperature and bias dependence of a TLF and its sensitivity to different gate electrodes allowing for a location
estimate~\cite{Ye202412,Li2018,Malcolm2020}. Additionally, utilizing detailed time
domain characterization, control of a single TLF has been demonstrated~\cite{Liu2018,Ye202407}, with the ability to reset or prepare the TLF in a particular state. We note that recently the cross-PSD has also
been used to gain location information~Ref.~\cite{Rojas-Arias2025}.

In this work, we analyze a prominent multi-level charge noise signal observed on a Si/SiGe device. The data were taken at multiple device configurations by sweeping different gate electrode voltage settings and recording a time series at each setting.  Our approach is to fit a number of factorial hidden Markov models (FHMMs) to the data and perform model selection to determine the most likely model for the data. To allow for the possibility of constituent charge fluctuators that consist of more than two states, we denote this more general case of an $n$-level fluctuator as an $n$LF. Once the model is determined, we obtain rate matrices for the $n$LF(s) as a function of the device sweep.  Interestingly, the time series displays multiple fluctuators at different time scales as well as a prominent three-level signal with equal level-spacing, reminiscent of Ref.~\cite{Uren1988}.

\section{Device and data}
\begin{figure}[ht]
\centering
{\includegraphics[width=0.24 \textwidth,trim = 10 0 0 0,clip]{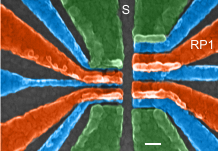}}
\caption{
\textbf{Device.}
False-color scanning electron micrograph of a device that is nominally identical to the one measured here. Screening gates, accumulation gates, plunger gates, and tunneling gates are in gray, green, red, and blue, respectively. The sensor quantum dot is tuned under plunger gate RP1. The white scale bar is 100 nm.
}
\label{fig:device}
\end{figure}

We measure a multi-dot device (Fig.~\ref{fig:device}) in a dilution refrigerator at a base temperature of $10~\si{mK}$. The device is fabricated in an overlapping gate architecture on an undoped Si/SiGe heterostructure with a natural Si quantum well of $8$-nm width approximately $50~\si{nm}$ below the semiconductor surface and with a $4$-nm thick Si capping layer. After the surface preparation, a thin $\sim$1 nm layer of SiO$_2$ forms on the Si capping layer. We apply a low voltage bias on the middle screening gate S to separate the left side and the right side (Fig.~\ref{fig:device}). We tune up a sensor quantum dot under plunger gate RP1 and the charge sensor is configured for rf reflectometry \cite{PhysRevApplied.13.024019}. We tune up the sensor dot in the Coulomb blockade regime and set the plunger gate voltage at the side of the Coulomb peak such that the measured conductance indicates fluctuations of dot electrochemical potentials.

The data consist of time series data sampled at 60Hz with a duration of approximately 1.14 hours, sweeping each voltage and temperature parameter independently while the unswept parameters sit at nominal values. See Table \ref{tab:params} for the specific parameter values.

\begin{table}[h]
    \begin{centering}
    \begin{tabular}{cccc}
    \toprule
    \textbf{Parameter} & \textbf{Name} & \textbf{Values} \\ \midrule
    P gate voltage & RP1 & 0.563V $\rightarrow$ 0.568V by 0.5mV\\
    S gate voltage & S & 0.148V $\rightarrow$ 0.152V by 0.5mV\\
    CS SD bias voltage & $V_{\mathrm{SD}}$ & (high, medium, low)\\
    MC temperature & $T_{\mathrm{MC}}$ & 100mK $\rightarrow$ 250mK by 30mK\\ \bottomrule
    \end{tabular}
    \end{centering}
   \caption{Parameters swept in the multi-level fluctuator dataset.}\label{tab:params}
\end{table}

\section{Data analysis methodologies and results}
\subsection{Detrending to account for drift and slow fluctuator removal}
\noindent
One challenge with this dataset is that, for certain time series at particular parameter sweep values, a significant amount of continuous drift (slow continuous-time wandering in signal) is present. This could be due to, for example, the collective effect of an ensemble of weaker $n$LFs or some continuous-time noise process. Since we focus on the discrete components of the noise fluctuations in our analysis, it is helpful to be able to ``subtract out'' this continuous drift and retain primarily the discrete fluctuations along with the non-drifting white noise background.

We use a sliding window-based Kernel Density Estimation (KDE) overlap comparison algorithm to track and determine drift. The algorithm is as follows. We slide a window of size 3,000 sample points (50 seconds) across each time series in steps of 50 sample points (0.42 seconds). At each step we compare the KDE of the first half of the window to the KDE of the second half.  We run an optimization routine to find the shift required to maximize the overlap of the two KDE's.
This distributional shift is what we use to track the drift. Since we are stepping through by 50 sample points we typically work with the linearly interpolated drift to apply directly to the raw time series.

\begin{figure*}
\centering
\begin{tabular}{cc}
  \includegraphics[width=80mm]{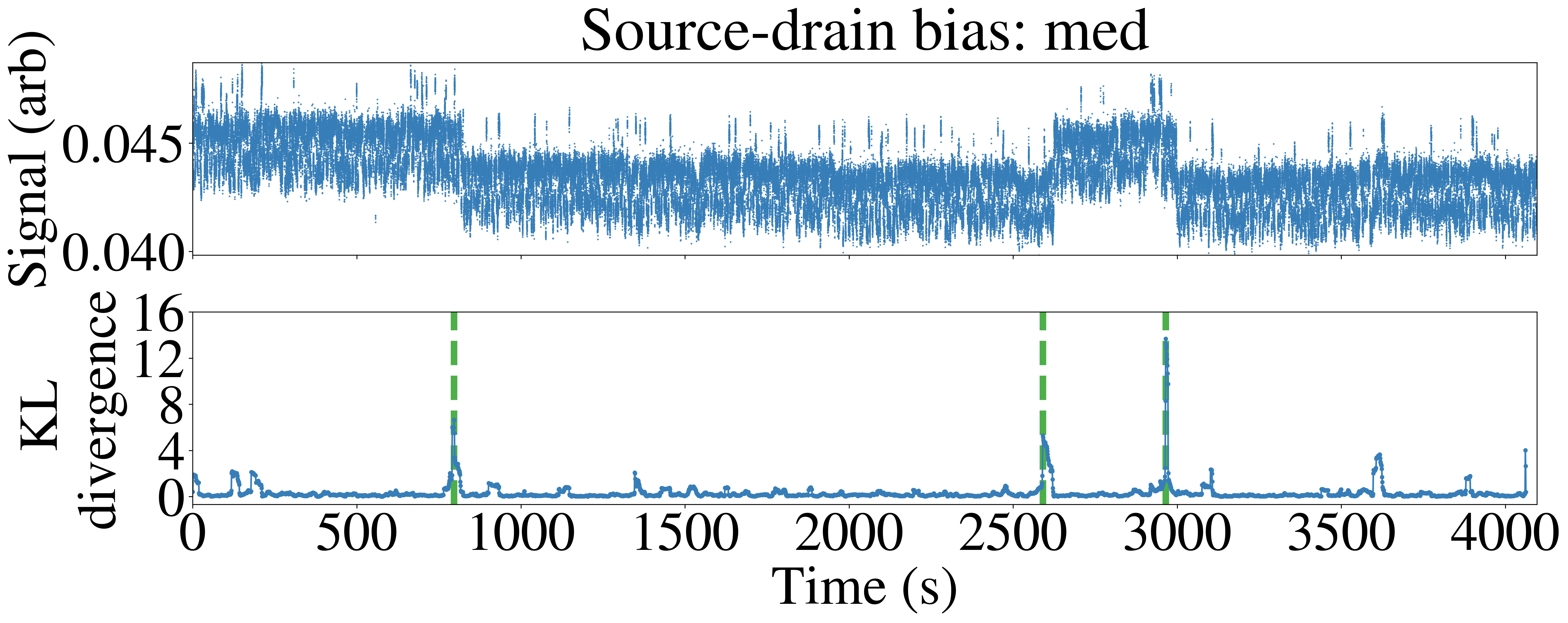} &
  \includegraphics[width=80mm]{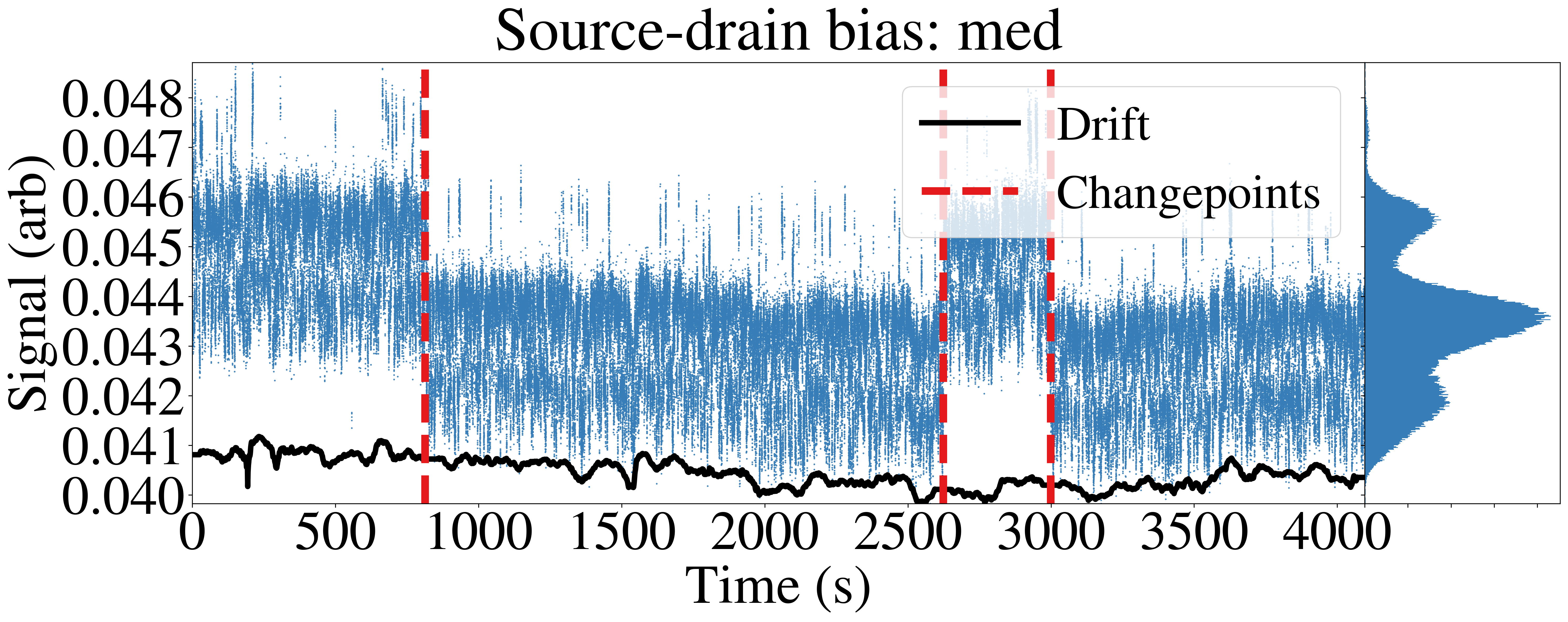}\\
(a) KLD method example &
(d) Raw timeseries with drift and jumps.  \\[6pt]
  \includegraphics[width=80mm]{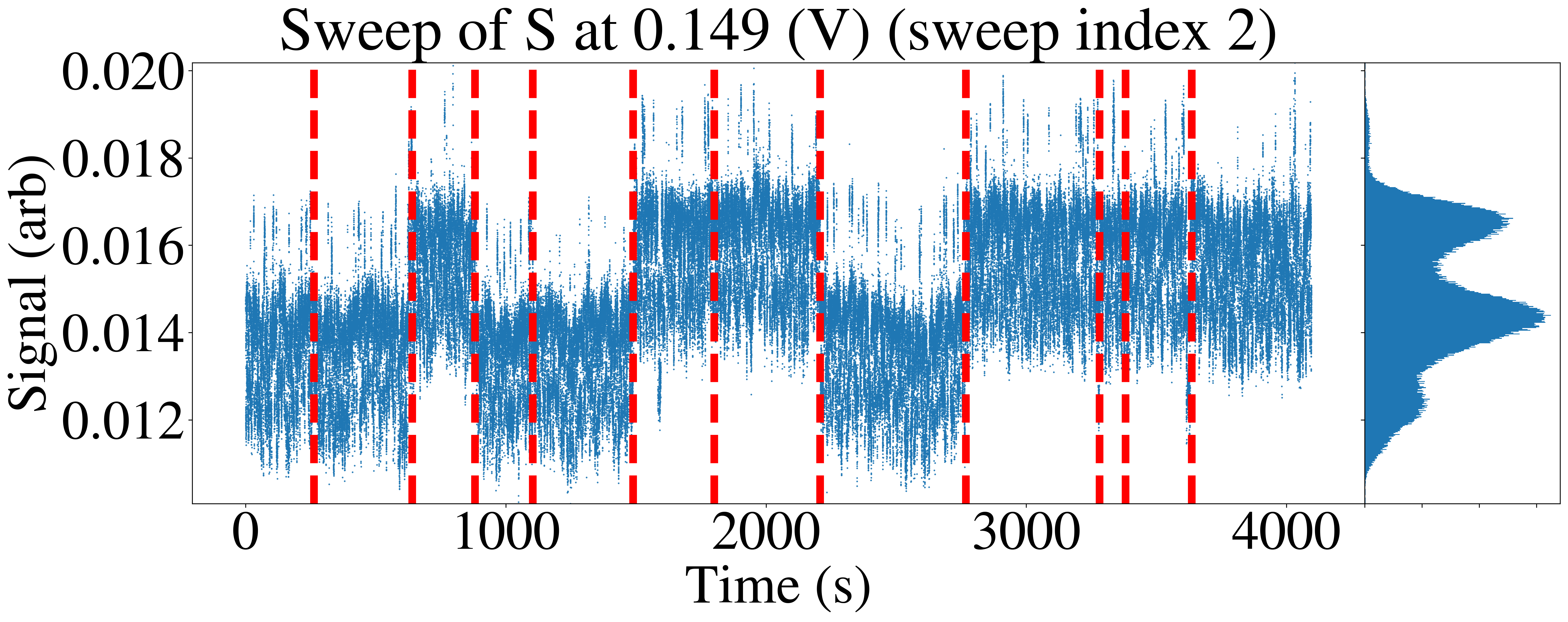} &
  \includegraphics[width=80mm]{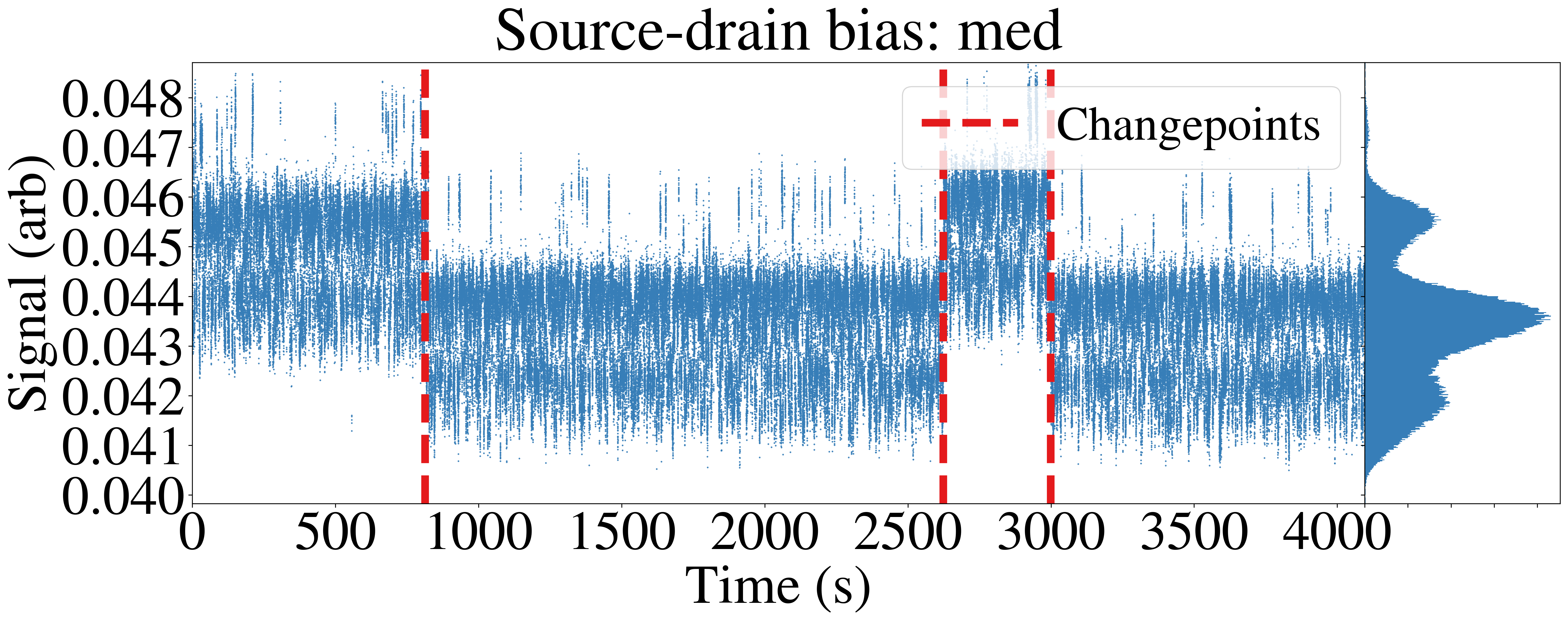}\\
(b) Raw time series with incorrect change point identification. &
(e) Drift-removed time series still containing jumps. \\[6pt]
  \includegraphics[width=80mm]{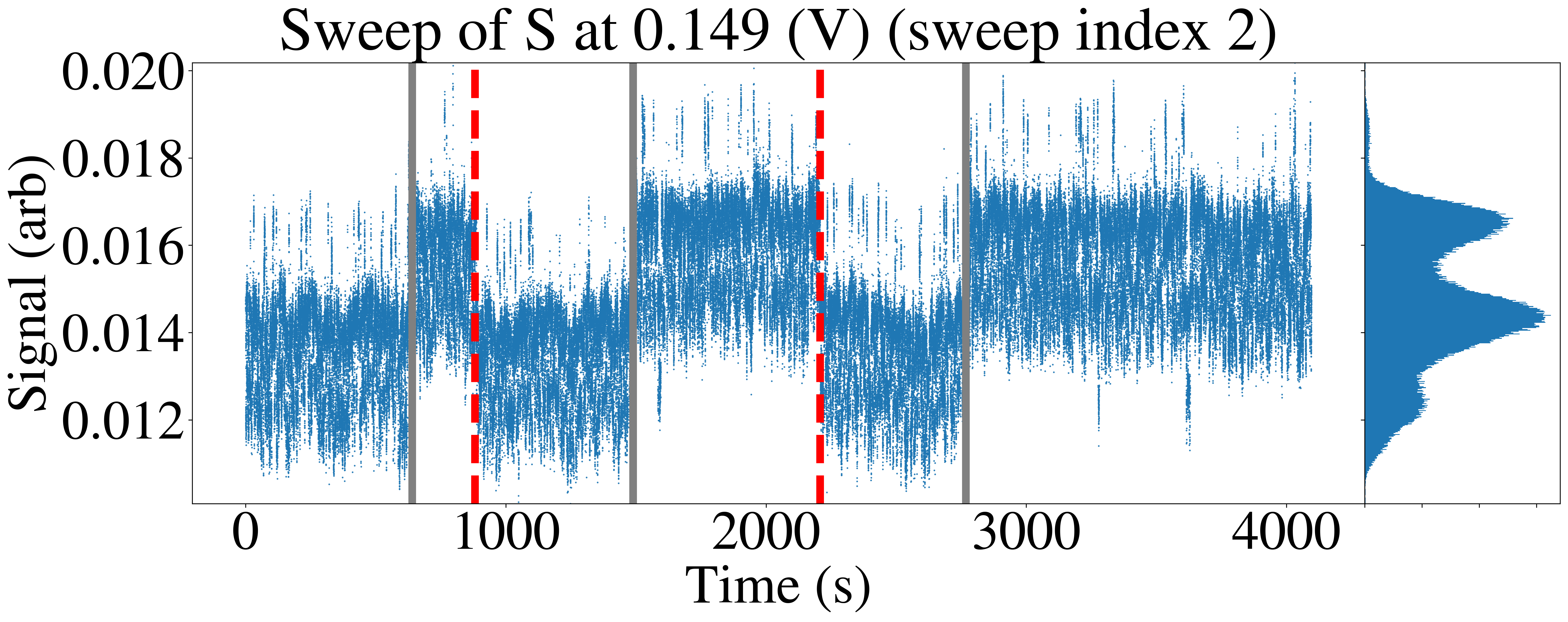} &
  \includegraphics[width=80mm]{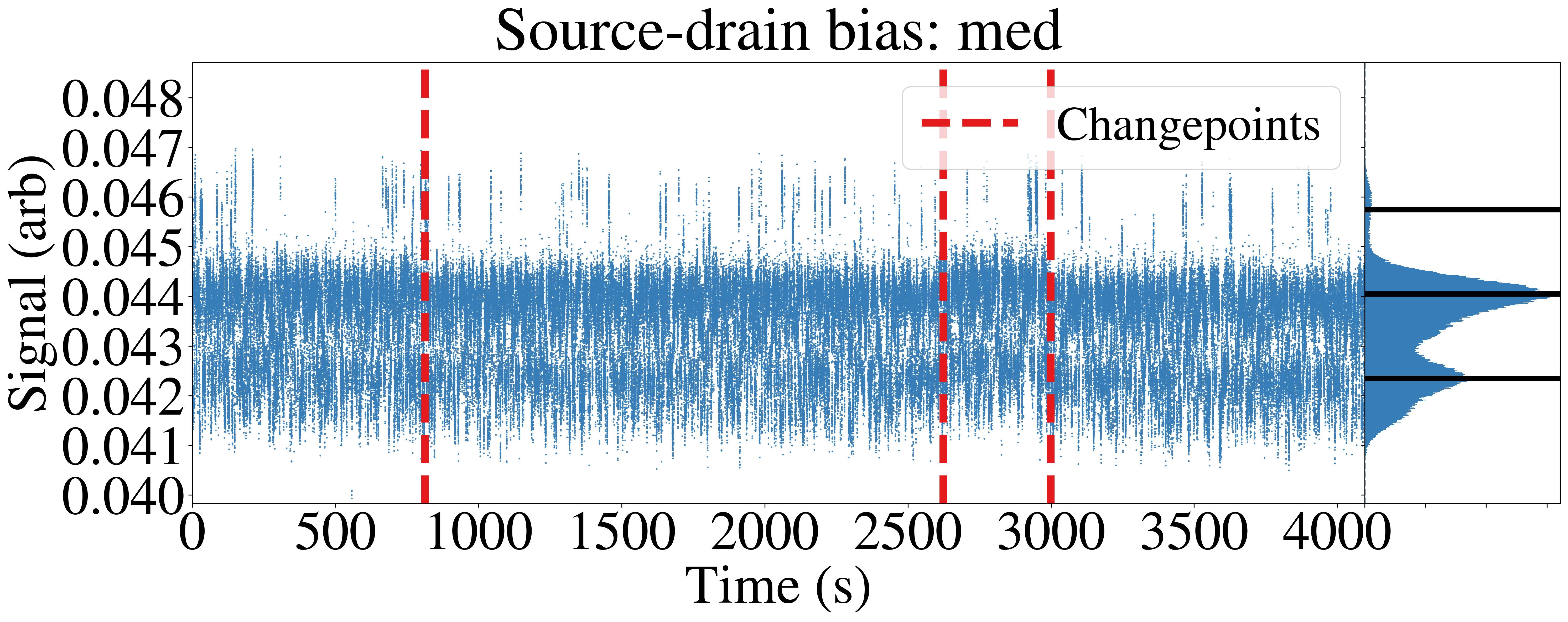}\\
(c) Raw time series with manually corrected change points. &
(f) Drift and jump removed timeseries. \\[6pt]
\end{tabular}
\caption{\textbf{Drift and change point removal.}  Figure (a) shows an example of change point detection using the KLD method described in the main text. A peak finding algorithm identifies large step changes indicated by the dashed lines. Figure (b) shows an example of change point detection failure, with dashed red lines indicating change points found by the algorithm. Figure (c) shows our manual pruning and adjustment corrections to (b), where we have used solid gray lines to mark the start of a higher-signal segment and dashed red lines to indicate the end of a higher-signal segment, which is also the start of a subsequent lower-signal segment. Figures (d), (e), and (f) show the process of detrending and removing the slow fluctuator going from (d) $\rightarrow$
(e) $\rightarrow$ (f).}
\label{fig:rmdrift_rmcps}
\end{figure*}

To assist in what would be an arduous manual change point detection, we implement a Kullback-Leibler Divergence (KLD) metric-based detrending algorithm utilizing Kernel Density Estimation (KDE), along with peak-finding to detect change points. At each step, the KDE of the first half of the window is compared to the KDE of the second half determining the KLD score. We use a peak-finding algorithm, with prominence and height parameters manually tuned to each sweep time series, to find peaks in the KLD time series, identifying change points, as shown in Fig.~\ref{fig:rmdrift_rmcps}(a).

Change point and drift detection are quite sensitive to parameter choices and do not always succeed. We make manual pruning and corrections when these automatically determined change points appear qualitatively incorrect. As an example, see Figs.~\ref{fig:rmdrift_rmcps}(b-c). The cases where misidentification is likely to arise are low amplitude and fast transitions of the problematic TLF.  This is most prevalent in the high mixing chamber temperature data.  Although some bias could be introduced from misidentification, we expect the impact of this to be low, since the rate of fluctuation of the residual fluctuators is quite high by comparison -- there would need to be quite a few misidentifications. We also show the process of detrending and change point removal in Figs.~\ref{fig:rmdrift_rmcps}(d-f).

\noindent
\subsection{FHMM fitting and model selection}
\noindent
We are interested in fitting the residual, post-processed time series fluctuator characteristics across parameter sweeps. This presents unique challenges, in that prominent features in one time series may disappear in another time series having a different parameter setting. In order to address this, we fit the data using only the three prominent peaks that appear widely across all but two time series, see the solid black lines in the histogram in Fig.~\ref{fig:rmdrift_rmcps}(f). The two time series that do not have three prominent levels are at the high voltage end of the RP1 sweeps, so we include in the analysis the RP1 time series up to, but not including, the final two. These peaks display an apparent equal level splitting, which we preserve in the fitting. 
We defer further comments on weight fixing.

\tikzset{
    level/.style = {
        ultra thick,
        black,
    },
    connect/.style = {
        dashed,
        ultra thick,
        black,
    },
    notice/.style = {
        draw,
        rectangle callout,
        callout relative pointer={#1}
    },
    label/.style = {
        text width=2cm
    }
}
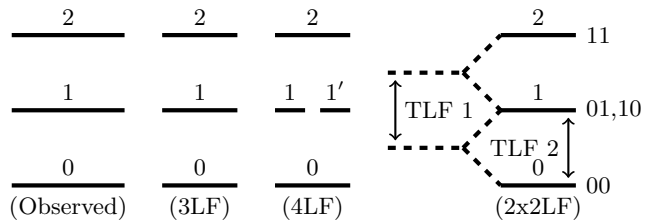
\begin{figure}
\centering
\begin{tikzpicture}

    \draw[level] (0,2) -- node[above] {$2$} (1.5,2);
    \draw[level] (0,1) -- node[above] {$1$} (1.5,1);
    \draw[level] (0,0) -- node[above] {$0$} node[below] {(Observed)} (1.5,0);

    \draw[level] (2,2) -- node[above] {$2$} (3,2);
    \draw[level] (2,1) -- node[above] {$1$} (3,1);
    \draw[level] (2,0) -- node[above] {$0$} node[below] {(3LF)} (3,0);

    \draw[level] (3.5,2) -- node[above] {$2$} (4.5,2);
    
    \draw[level] (3.5,1) -- node[above] {$1$} (3.9,1);
    \draw[level] (4.1,1) -- node[above] {$1'$} (4.5,1);
    
    \draw[level] (3.5,0) -- node[above] {$0$} node[below] {(4LF)} (4.5,0);

    \draw[connect] (5,1.5) -- (6,1.5);
    \draw[connect] (5,0.5) -- (6,0.5);

    \draw[<->, thick] (5.1,0.6) -- node[right] {TLF 1} (5.1,1.4);

    \draw[connect] (6,1.5) -- (6.5,2) (6,1.5) -- (6.5,1);
    \draw[connect] (6,0.5) -- (6.5,1) (6,0.5) -- (6.5,0);
    
    \draw[level] (6.5,2) -- node[above] {$2$} (7.5,2) node[right] {11};
    \draw[level] (6.5,1) -- node[above] {$1$} (7.5,1) node[right] {01,10};
    \draw[<->, thick] (7.4,0.1) -- node[left] {TLF 2} (7.4,0.9);
    \draw[level] (6.5,0) -- node[above] {$0$} node[below] {(2x2LF)} (7.5,0) node[right] {00};
\end{tikzpicture}
\caption{\textbf{Models.} We show the breakdown of fitting different models to the 3 prominent observed levels in the data (left). The levels for the 3LF and 4LF models are set as shown. For the 2x2LF model, we show how the levels add indicating the final levels on the far right.}
\label{fig:levels}
\end{figure}

We fit three different statistical models to the post-processed data.
These models are shown schematically in Fig.~\ref{fig:levels}.
The 2x2LF model takes two independent two-level fluctuators such that
their level spacings add to create the three observed levels.  We fit a 3LF model, which is a single three-level model.  Finally, we fit a 4LF model where we fix the middle levels to be the same, to reproduce the three observed levels. We compare these models using the evidence ratio
\cite{Nielsen2021, Albrecht2023} to determine the best model.  The evidence ratio
is given by
\begin{equation}
e = 2 \, \frac{\ln \mathcal{L}_i - \ln \mathcal{L}_{j}}{N_{i} - N_{j}}
\end{equation}
where $\ln \mathcal{L}$ is the log likelihood, $N$ is the number of parameters of the model, $i$ refers to the larger model, and $j$ refers to the smaller model.
If $e < 1$, there is no evidence against the smaller model. If $1 < e < 2$, there is weak evidence against the smaller model. Finally, if $e > 2$ there is strong evidence for the larger model over the smaller model. For most of the sweep values we do not find a significant difference in performance between the different models. We show the model comparisons for all swept gate electrodes in Fig. \ref{fig:model_comparison}. However, for $\textrm{RP1}=0.567 \textrm{V}$, we find that the 4LF model outperforms the 2x2LF model and vastly outperforms the 3LF model, as shown in Fig.~\ref{fig:RP1_MC_and_4LF_rates}(a). This is an indication of a conditional dependence between two 2LFs, wherein the transition rates between the states of one fluctuator depend on the state of the other fluctuator. We expand upon this further in Section \ref{subsec:conditional_rate_analysis}.

\begin{figure*}[ht!]
\centering
\begin{tabular}{cc}
  \includegraphics[width=75mm]{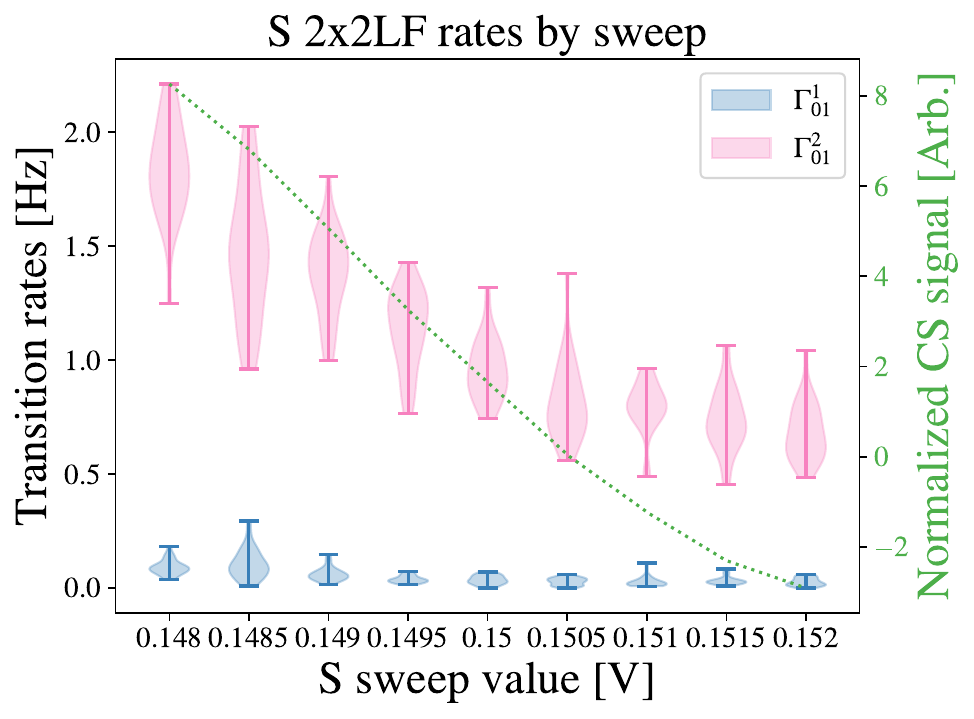} & 
  \includegraphics[width=75mm]{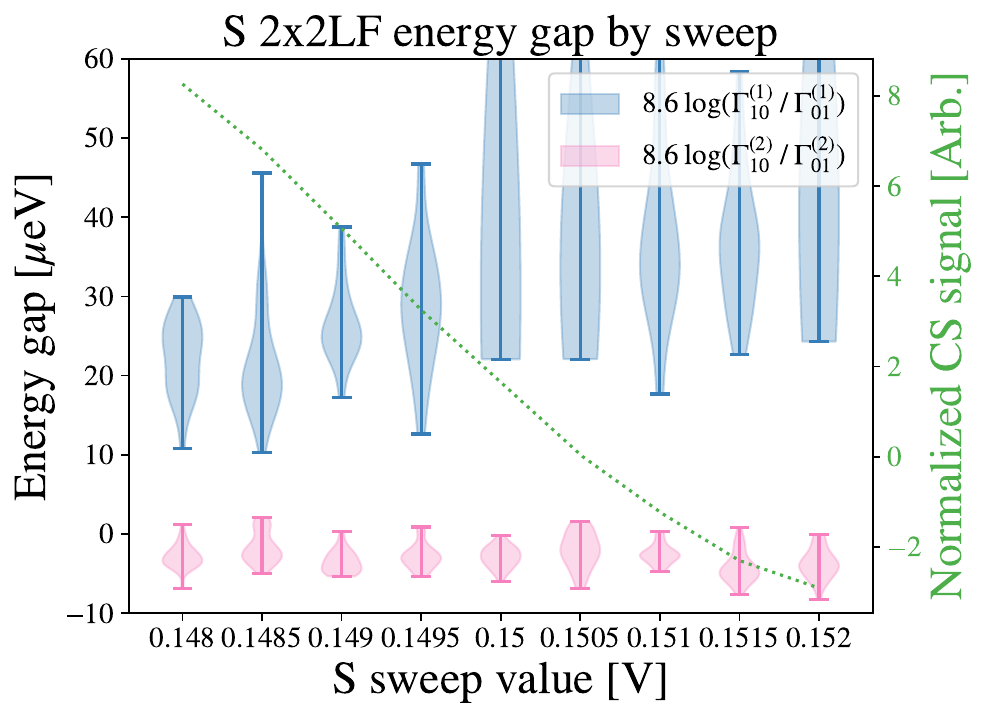}\\
  (a) & (b)  \\[6pt]
  \includegraphics[width=75mm]{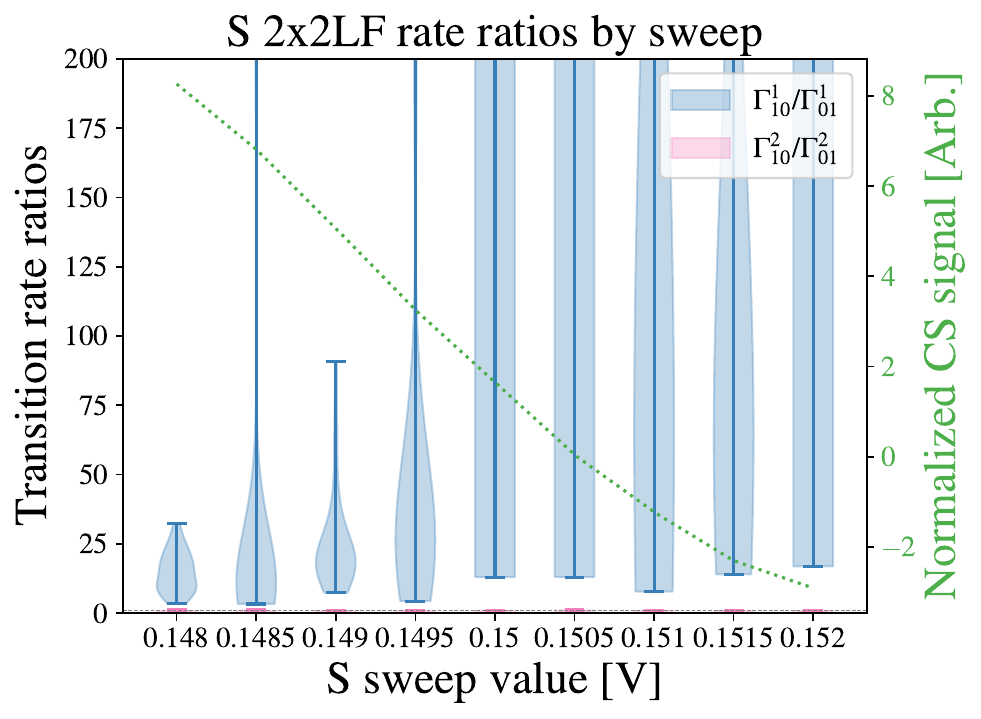} &
  \includegraphics[width=75mm]{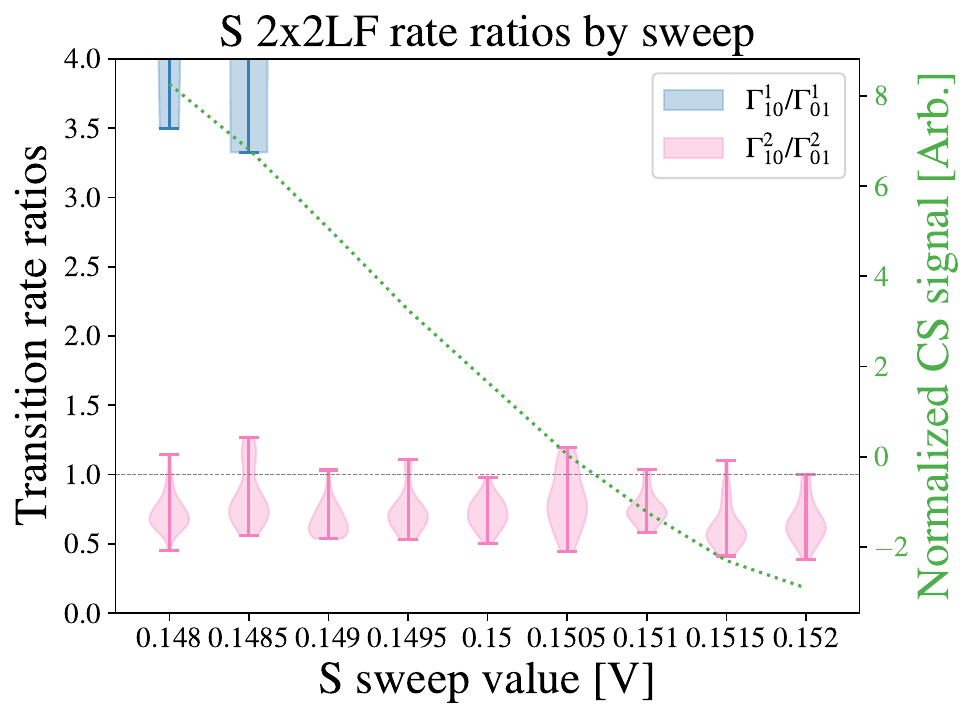} \\
  (c) Focusing on $\Gamma^{1}$  & (d) Focusing on $\Gamma^{2}$  \\[6pt]
\end{tabular}
\caption{\textbf{S sweep FHMM rate fits for the 2x2LF model.}  The FHMM
fit rates for each sweep value.}
\label{fig:S_rates}
\end{figure*}

\begin{figure*}[ht!]
\centering
\begin{tabular}{cc}
  \includegraphics[width=75mm]{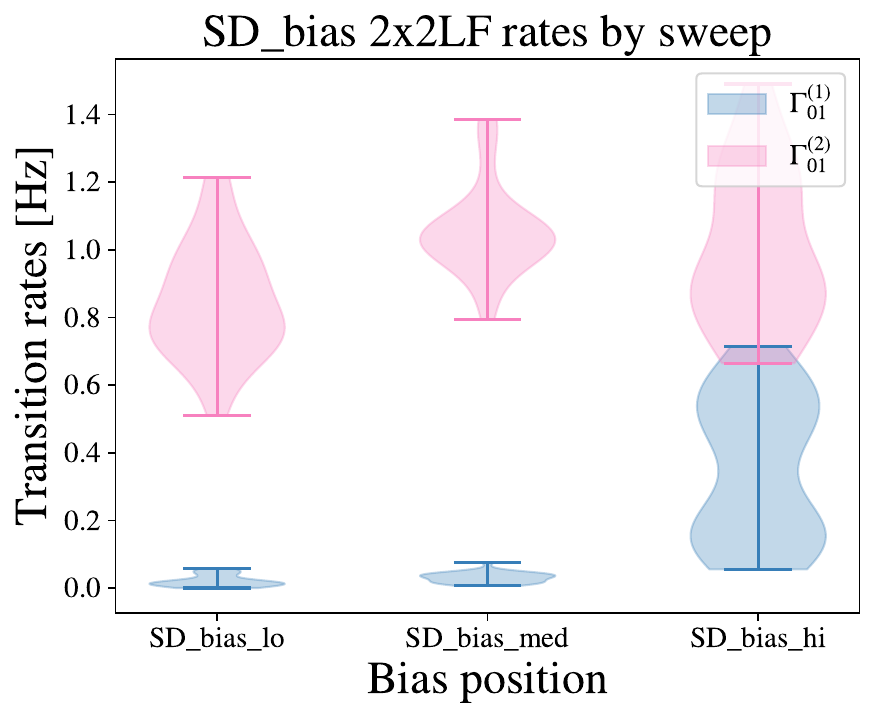} & 
  \includegraphics[width=75mm]{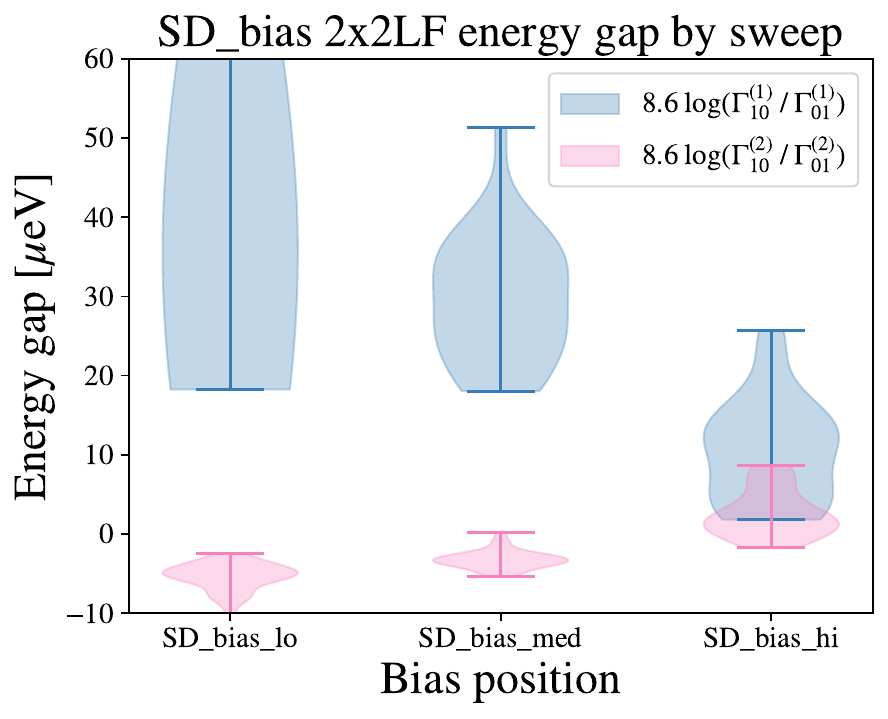}\\
  (a) & (b)  \\[6pt]
  \includegraphics[width=75mm]{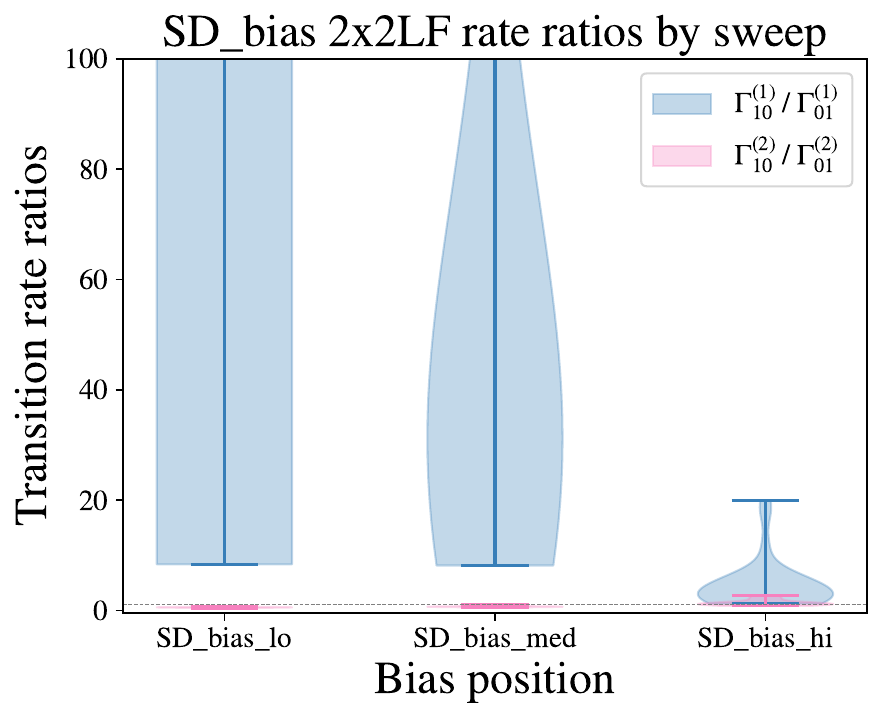} &
  \includegraphics[width=75mm]{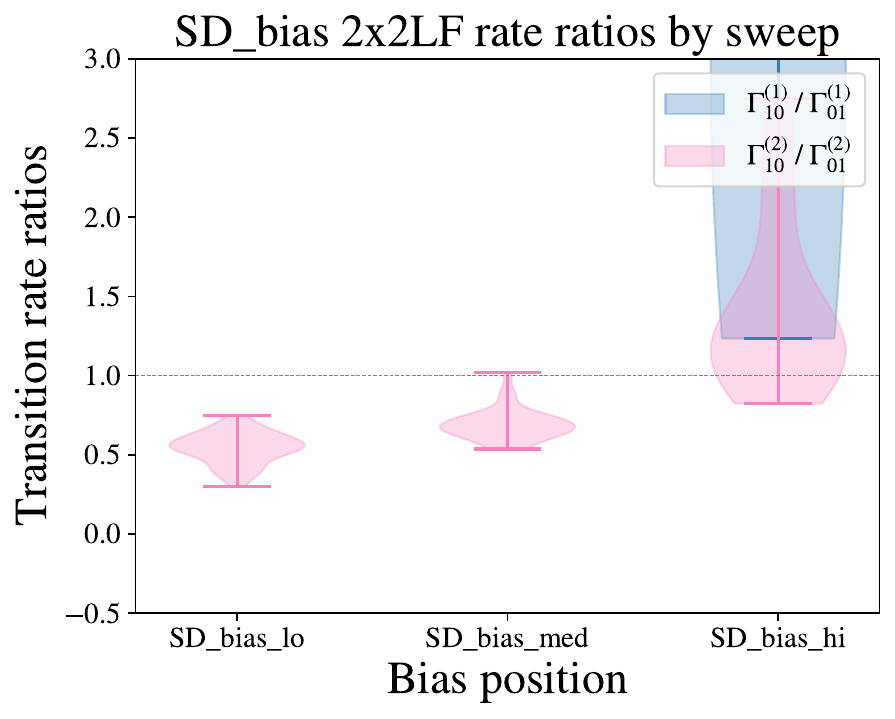} \\
  (c) Focusing on $\Gamma^{1}$  & (d) Focusing on $\Gamma^{2}$  \\[6pt]
\end{tabular}
\caption{\textbf{SD bias sweep FHMM rate fits for the 2x2LF model.}  The
FHMM fit rates for each sweep value.}
\label{fig:SD_bias_rates}
\end{figure*}

\begin{figure*}[ht!]
\centering
\begin{tabular}{cc}
  \includegraphics[width=75mm]{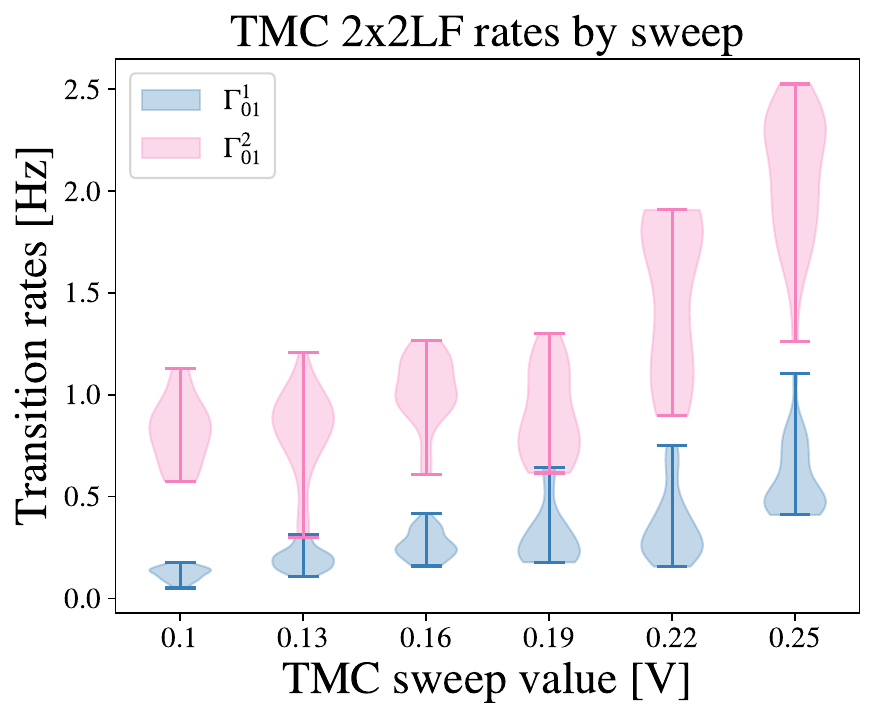} & 
  \includegraphics[width=75mm]{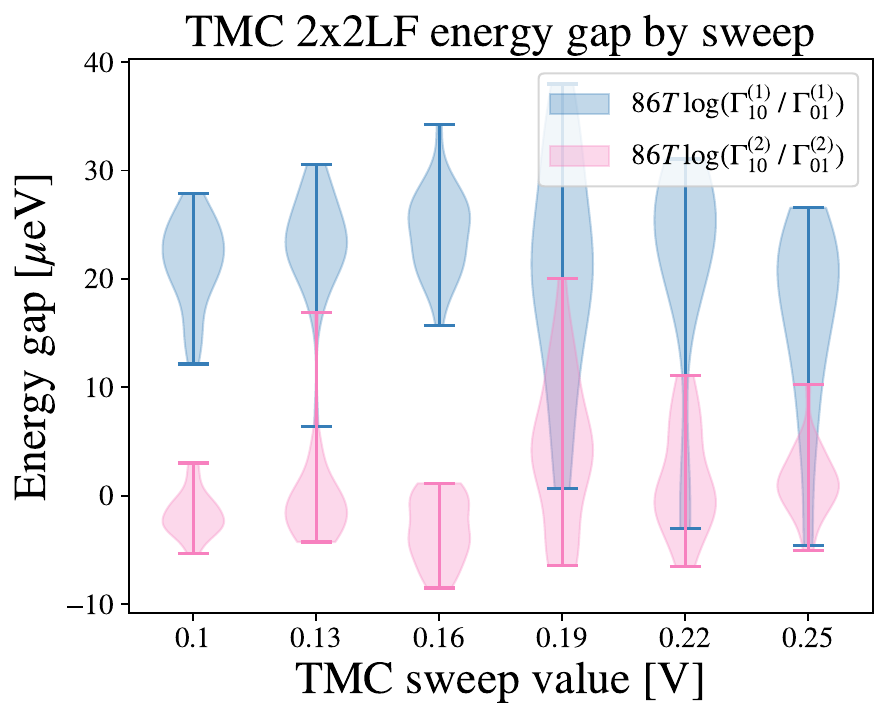}\\
  (a) & (b)  \\[6pt]
  \includegraphics[width=75mm]{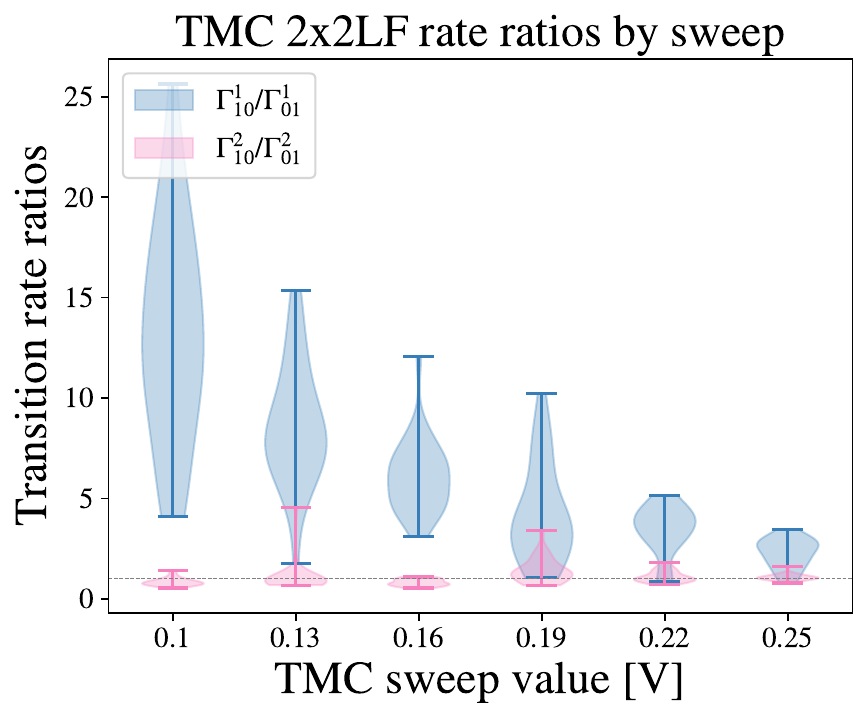} &
  \includegraphics[width=75mm]{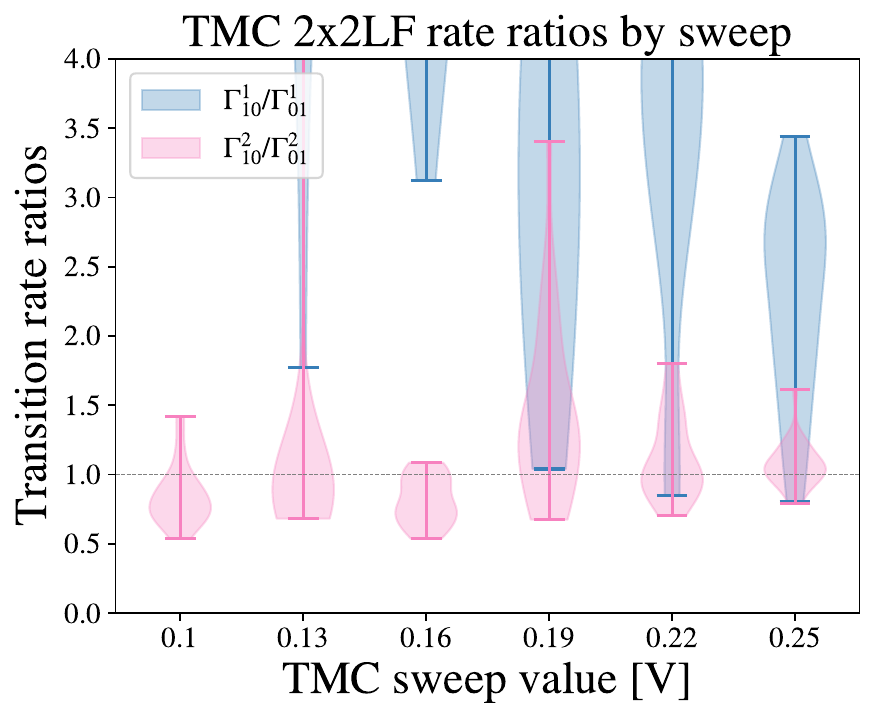} \\
  (c) Focusing on $\Gamma^{1}$  & (d) Focusing on $\Gamma^{2}$  \\[6pt]
\end{tabular}
\caption{\textbf{TMC sweep FHMM rate fits for the 2x2LF model.}  The FHMM
fit rates for each sweep value.}
\label{fig:TMC_rates}
\end{figure*}

For the majority of the parameter sweeps, we report the results of the 2x2LF model.  This model is the simplest and performs as well as the other models, when comparing the likelihoods and Akaike Information Criterion (AIC) scores.  The transition rates, the ratios of forward and reverse rates, as well as the energy gap for all sweeps are shown in
Figs.~\ref{fig:RP1_rates},\ref{fig:S_rates},\ref{fig:SD_bias_rates},\ref{fig:TMC_rates}.
We plot the $30$ bootstrap samples for each rate represented by a violin plot.  In Section
\ref{sec:phenomenological_model} we extract physical estimates from the sweep dependence
shown in these rates.

\subsubsection{Comments on the level tracking problem and weight fixing}
\noindent
There are certainly more than three levels in the post-processed data. For example, in Fig.~\ref{fig:rmdrift_rmcps}(f), there are clearly side-lobes visible on the three identified peaks, corresponding to additional levels.  However, these disappear in the noise in other sweeps, confounding both pre-fit weight fixing of these levels for tracking purposes and (in the case of allowing the weights to vary) post-fit identification between models of different time series. As such, we do not treat these less prominent levels that cannot be tracked across the majority of the datasets, focusing instead on three prominent, well-separated levels. In particular, we are interested in capturing the sparsely-populated, highest-signal level of the dataset, as shown by the top solid black line in the histogram of Figure \ref{fig:rmdrift_rmcps}(f). If we allow the weights to vary freely, we found through trial-and-error that there are some time series containing less well-separated, though more densely-sampled, level shifts to which the weights of the model will (frustratingly) fit instead. These new levels are not the desired, well-separated levels and are not robustly captured across all datasets, resulting in what we call a \emph{level tracking problem}. We address this problem by fixing the weights to track three prominent levels in the data that are present in all sweeps analyzed. The \emph{level tracking problem} is a distinct problem as compared to the issue of identifying fluctuators across different model fits where the independent models do not restrict which fluctuators belong to which ordering of parameters, which we call the \emph{identification problem}, to be discussed next.

\subsubsection{A note on the identification and embedding problems}
\noindent
In the \emph{identification problem}, we attempt to match models across multiple bootstraps as well as across multiple sweeps. Since the FHMM algorithm doesn't fix an ordering to the fluctuators, it is possible in the 2x2LF case for the identification of one TLF for one fit to switch places with the next fit. This creates a problem with identification, when attempting to aggregate results across multiple fits.  Additionally, for the 4LF model, since the levels are degenerate, one fitted model could have a swapped
association of the rates to and from the intermediate levels ($1$ and $2$ in Fig.~\ref{fig:levels}).  Hence, for the 2x2LF model we might need to swap the fluctuators and for the 4LF model we might need to swap the levels. We identify models for the 2x2LF case by keeping $\Gamma^{1}_{01} < \Gamma_{01}^{2}$, and for the 4LF case we keep $\Gamma_{01} < \Gamma_{02}$, where $\Gamma_{ij}^{\alpha}$ is the rate for transitioning from state $j$ to state $i$ of fluctuator $\alpha$. If the rates shift from one 2LF to the other (2x2LF) or there's a bias flip (4LF), this may lead to a misidentification.

As a final technical detail, we must address the \emph{embedding problem} (see for example
\cite{Johansen1973, Davies2010, Baake2020}) in order to properly extract rates from our models and continue with our modeling of the system. As formulated in the \emph{embedding problem}, we must check whether the transition matrix $M$ may be generated from a continuous-time Markov chain:
\begin{equation}
M = e^{\Gamma t} \quad .
\end{equation}
We employ some simple tests on the 2x2LF \cite{Kingman1962} and 3LF
\cite{Johansen1974} models for embeddability, as well as for uniqueness of the
principal logarithm \cite{Davies2010}.  For the 4LF model, we appeal to the algorithm of \cite{Casanellas2023}.
If a Markov matrix $M$ cannot be embedded, we use the diagonal adjustment algorithm of Refs.~\cite{Kreinin2001, Davies2010} to approximate the associated rate matrix $\Gamma \xrightarrow{d.a.} \Gamma'$ (restricting to cases where $\Gamma$ has real entries).  We then calculate the associated transition matrix $M' = e^{\Gamma't}$ and check that the log likelihood has not significantly changed.
\\
\subsection{Conditional rate analysis}
\label{subsec:conditional_rate_analysis}
\begin{figure*}[ht!]
\centering
\begin{tabular}{cc}
  \includegraphics[width=75mm]{Figs/RP1_model_comparisons.pdf} &   \includegraphics[width=75mm]{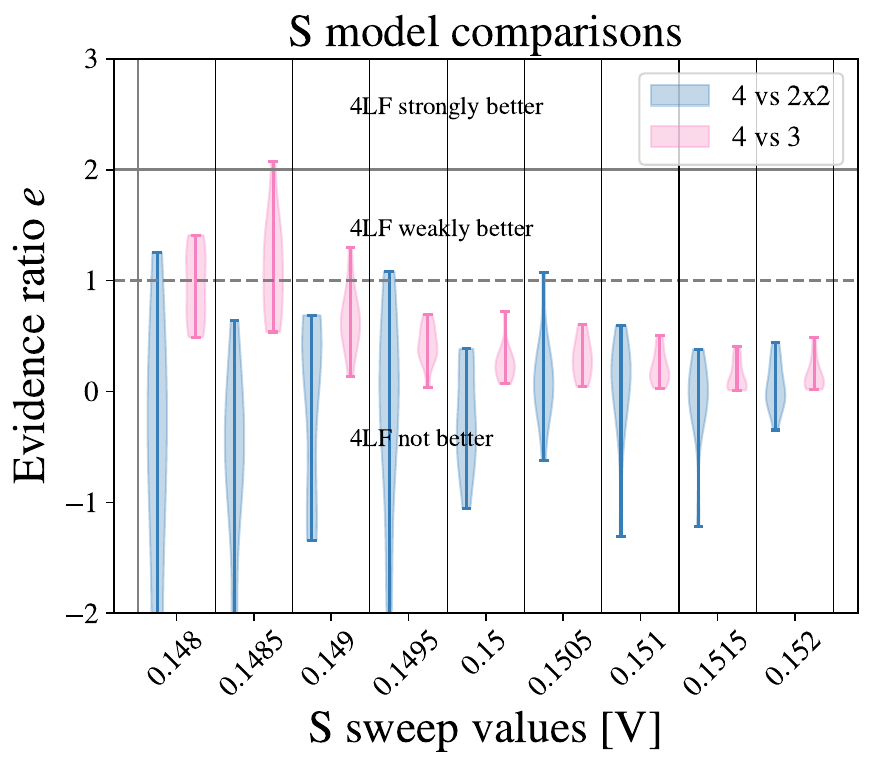} \\
(a)  & (b)  \\[6pt]
 \includegraphics[width=75mm]{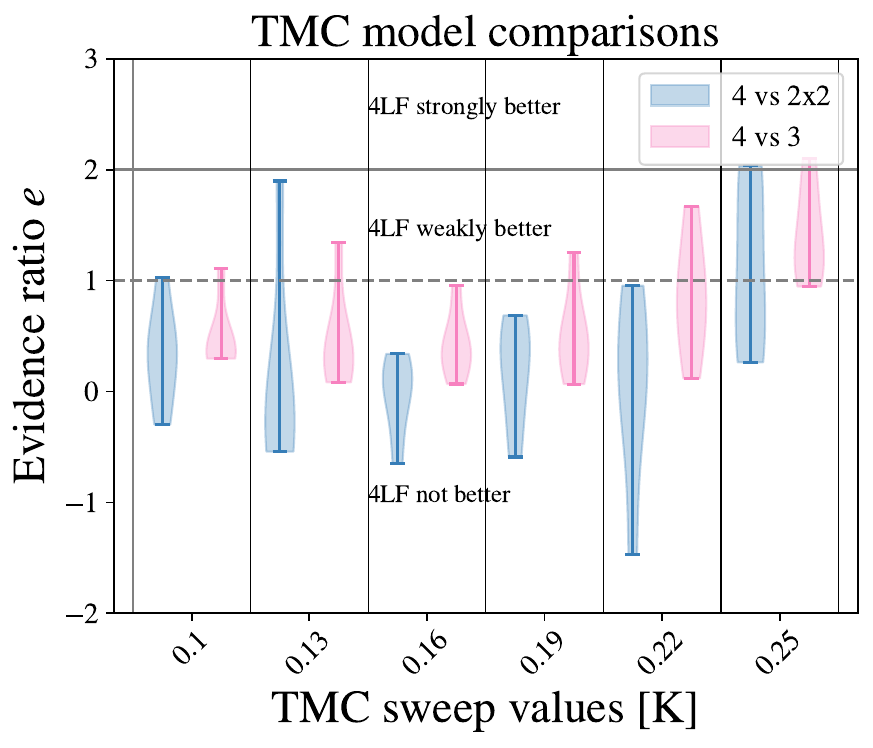} &   \includegraphics[width=75mm]{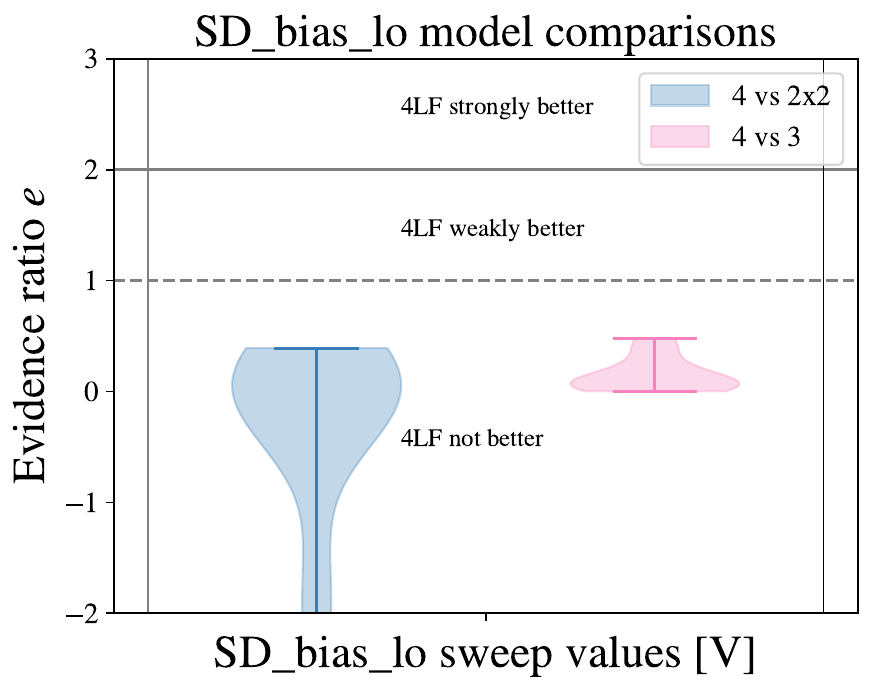} \\
(c)  & (d)  \\[6pt]
 \includegraphics[width=75mm]{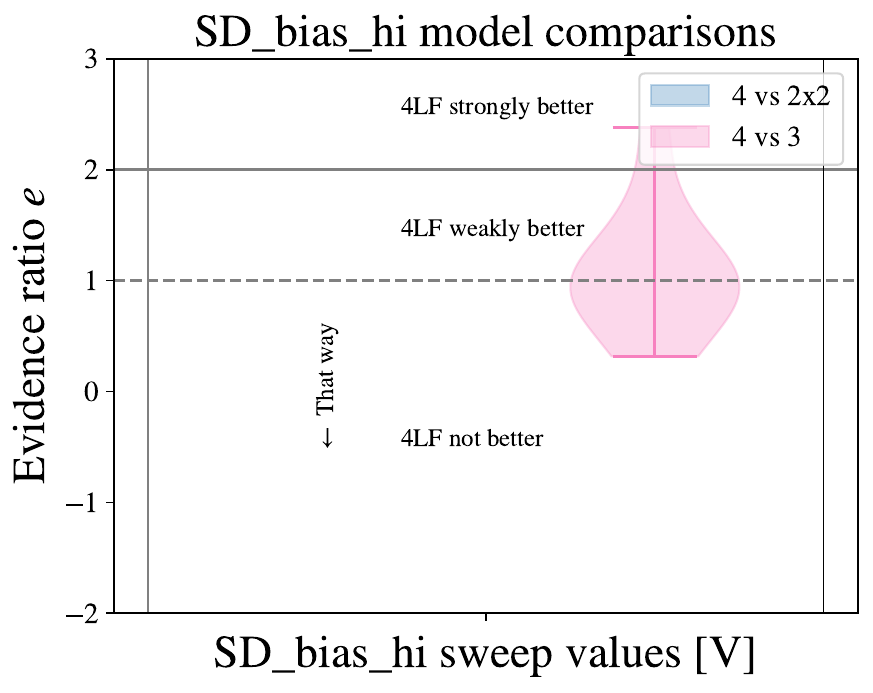} &   \includegraphics[width=75mm]{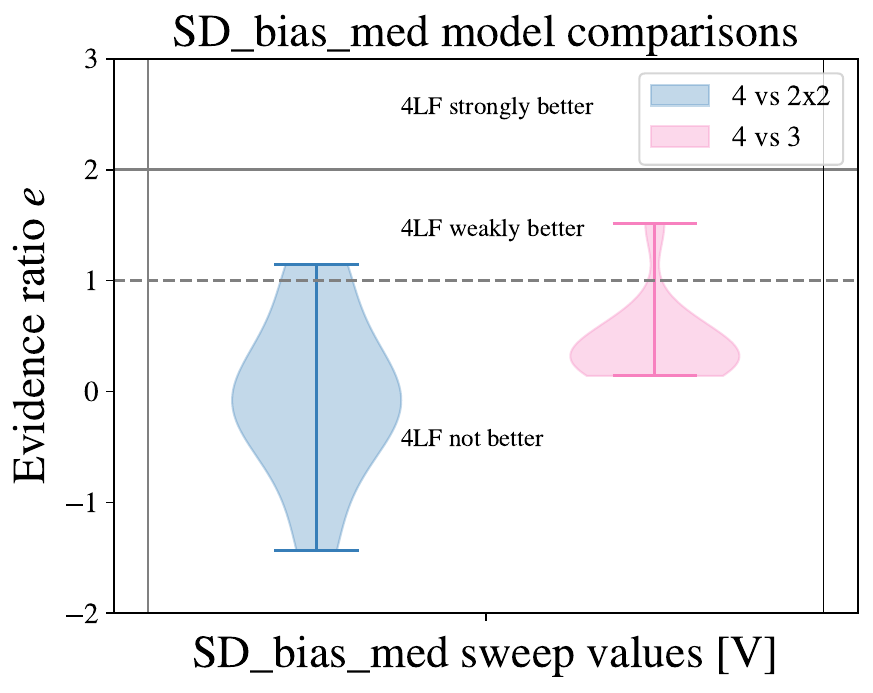} \\
(e)  & (f)  \\[6pt]
\end{tabular}
\caption{\textbf{Model comparisons.}  Here we show bootstrapped
evidence ratio model comparisons for different sweep values and
configurations.  For example, in figure (a) we can see the 4LF model is
not better until we reach RP1$=0.567$ V. We have grayed out rate
calculations at sweep values where there is significant aliasing in the
time series.}
\label{fig:model_comparison}
\end{figure*}

\noindent
In the case of $\textrm{RP1} = 0.567$ V device configuration, we find that the 4LF model clearly outperforms the 2x2LF model, see Fig.~\ref{fig:model_comparison}.  We are interested in the violation away from a 2x2LF model that is supported by this better-fitting 4LF model. As such, we start with representing the 2x2LF model within a larger 4LF model.
\begin{figure*}[ht!]
\centering
\begin{tabular}{cc}
  \includegraphics[width=75mm]{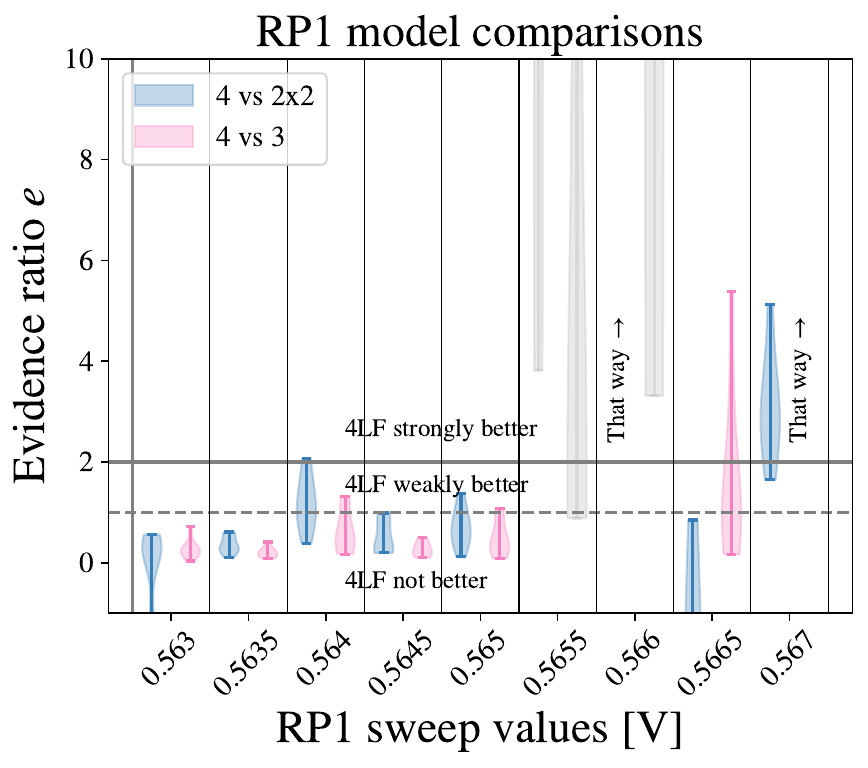} &
  \includegraphics[width=75mm]{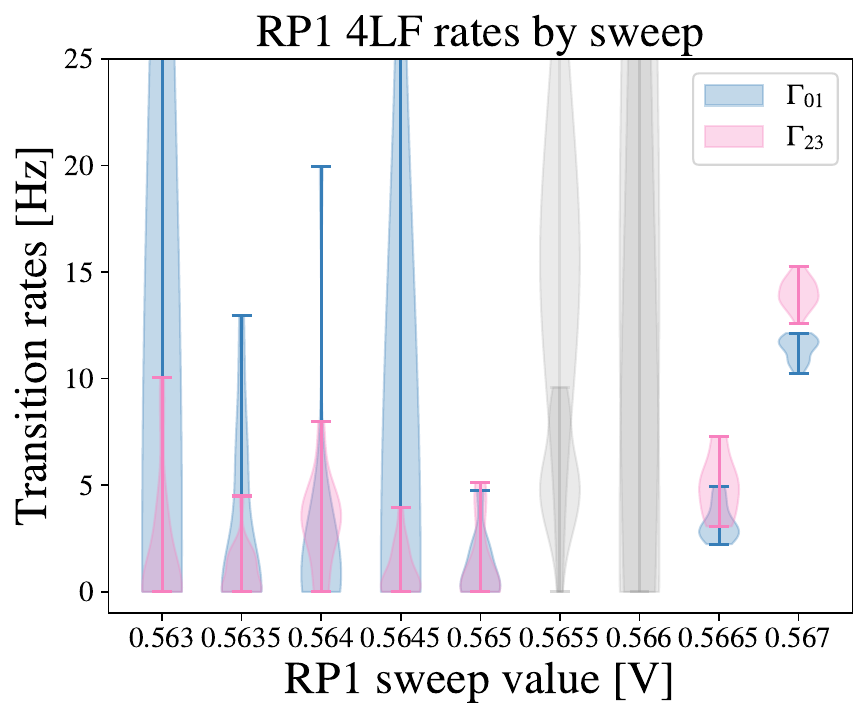} \\ 
  (a) Model comparison: 4LF vs 3LF and 4LF vs 2x2LF. & (b) $\Gamma_{01} = \Gamma_{23}$ for embedded independent 2x2LF.  \\[6pt]
\end{tabular}
\caption{\textbf{RP1 model comparison and FHMM rate fits for the 4LF model.} In (a), we show the model comparisons as a function of RP1, indicating that the 4LF model becomes better ($e > 2$) at the largest RP1 value. In (b), we display specific 4LF FHMM fit transition rate elements, $\Gamma_{01}$ and $\Gamma_{23}$, for each sweep value. These elements should be the same under a 2x2LF model. We can see a clear deviation at sweep value RP1$=0.567V$. In both figures we have grayed out rate calculations at sweep values where there is significant aliasing in the time series.}
\label{fig:RP1_MC_and_4LF_rates}
\end{figure*}
The equation of motion for our 4LF system follows the general form
\begin{equation}
\dot{\vec{P}} = \Gamma \cdot \vec{P}
\end{equation}
Considering two independent processes $P^1$ and $P^2$, we write down the equation of
motion for the joint probability $P_{(i,a)} = P_{i}^{1} P_{a}^{2}$ as
\begin{eqnarray}
\dot{P}_{(i,a)} & = & \dot{P}_{i}^{1} P_{a}^{2} + P_{i}^{1} \dot{P}_{a}^{2} \nonumber \\
                & = & \Gamma_{ij}^{1} P_{j}^{1} P_{a}^{2}
                  + \Gamma_{ab}^{2} P_{i}^{1} P_{b}^{2} \nonumber \\
                & = & \Gamma_{(i,a)(j,b)} P_{j,b} \quad ,
\end{eqnarray}
where for two independent 2LFs we have that
\begin{equation}
\Gamma^{1} = \left[
\begin{matrix}
-\gamma_{10} & \gamma_{01} \\
\gamma_{10} & -\gamma_{01} \\
\end{matrix}
\right]\quad {\rm and}
\quad
\Gamma^{2} = \left[
\begin{matrix}
-\lambda_{10} & \lambda_{01} \\
\lambda_{10} & -\lambda_{01} \\
\end{matrix}
\right] \quad ,
\end{equation}
implying that the 4LF rate matrix takes the form
\begin{eqnarray}
& \Gamma = \Gamma^{1} \otimes \mathlarger{\mathbb{1}} + \mathlarger{\mathbb{1}} \otimes \Gamma^{2} = \nonumber \\
&
\begin{tikzpicture}[baseline=-0.5ex]
  \matrix(m)[matrix of math nodes, nodes in empty cells,column sep=1em,row sep=1em,{left delimiter=[},{right delimiter=]}, nodes={text width={39}, align=center, inner sep=2pt},ampersand replacement=\&]{
    -\lambda_{10} - \gamma_{10} \& \lambda_{01} \& \gamma_{01} \& 0 \\
    \lambda_{10} \& -\lambda_{01} - \gamma_{10} \& 0 \& \gamma_{01} \\
    \gamma_{10} \& 0 \& -\gamma_{01} - \lambda_{10} \& \lambda_{01} \\
    0 \& \gamma_{10} \& \lambda_{10} \& -\gamma_{01} - \lambda_{01} \\
  };
  \draw[red](m-1-2.north-|m-1-2.west) rectangle (m-1-2.east|-m-1-2.south);
  \draw[red](m-3-4.north-|m-3-4.west) rectangle (m-3-4.east|-m-3-4.south);
  \draw[red,dashed](m-2-1.north-|m-2-1.west) rectangle (m-2-1.east|-m-2-1.south);
  \draw[red,dashed](m-4-3.north-|m-4-3.west) rectangle (m-4-3.east|-m-4-3.south);

  \draw[blue](m-1-3.north-|m-1-3.west) rectangle (m-1-3.east|-m-1-3.south);
  \draw[blue](m-2-4.north-|m-2-4.west) rectangle (m-2-4.east|-m-2-4.south);
  \draw[blue,dashed](m-3-1.north-|m-3-1.west) rectangle (m-3-1.east|-m-3-1.south);
  \draw[blue,dashed](m-4-2.north-|m-4-2.west) rectangle (m-4-2.east|-m-4-2.south);
\end{tikzpicture},
\end{eqnarray}
where we have placed boxes around matrix elements that would have the same value.
If there is a deviation from the 2x2LF model, we expect to see a corresponding deviation
in the constraints of the corresponding 4LF rate matrix. Averaging the bootstrap fit data for the $\textrm{RP1} = 0.567$ V voltage configuration, we have the following rate matrix for the 4LF model, including the standard error of the mean:

\begin{equation}
\widehat{\Gamma} = 
\begin{tikzpicture}[baseline=-0.5ex]
  \matrix(m)[matrix of math nodes, nodes in empty cells,column sep=0.7em,row sep=1em,{left delimiter=[},{right delimiter=]}, nodes={text width={36}, align=right, inner sep=1pt}]{
         -10.70(7) & 11.4(1) & 1.38(6) & 0.017(5) \\
         10.03(8) & -11.6(1) & 0.07(2) & 0.16(6)  \\
         0.66(3) & 0.14(2) & -13.8(1) & 14.0(1)   \\
         0.015(5) & 0.06(2) & 12.3(1) & -14.2(1)   \\
  };
  \draw[red](m-1-2.north-|m-1-2.west) rectangle (m-1-2.east|-m-1-2.south);
  \draw[red](m-3-4.north-|m-3-4.west) rectangle (m-3-4.east|-m-3-4.south);
  \draw[red,dashed](m-2-1.north-|m-2-1.west) rectangle (m-2-1.east|-m-2-1.south);
  \draw[red,dashed](m-4-3.north-|m-4-3.west) rectangle (m-4-3.east|-m-4-3.south);

  \draw[blue](m-1-3.north-|m-1-3.west) rectangle (m-1-3.east|-m-1-3.south);
  \draw[blue](m-2-4.north-|m-2-4.west) rectangle (m-2-4.east|-m-2-4.south);
  \draw[blue,dashed](m-3-1.north-|m-3-1.west) rectangle (m-3-1.east|-m-3-1.south);
  \draw[blue,dashed](m-4-2.north-|m-4-2.west) rectangle (m-4-2.east|-m-4-2.south);
\end{tikzpicture}
\end{equation}
The model fit is structurally similar to that expected from embedding two independent TLF rate matrices -- the anti-diagonal elements are close to zero, and the elements that should be identical are on par with one-another. The blue boxes look to correspond to a slow TLF, while the red boxes corresponding to a fast TLF. However, there is clear deviation from the strict embedding -- the blue solid boxes, which would be identical for embedded 2x2LF rate matrices, are statistically different (as are the blue dashed, red solid, and red dashed). This is shown graphically in Fig.~\ref{fig:RP1_MC_and_4LF_rates}(b), where we can see a clear indication that $\Gamma_{01} \neq \Gamma_{23}$
, as we go to high RP1 voltage.
Thus, between the two TLFs, there is evidence for a \emph{conditional switching dependence at a particular voltage} configuration.

One possible explanation for this conditional dependence is that the two charge fluctuators are physically nearby, such that the energy bias of one fluctuator is perturbed by the state of the other fluctuator. Within this physical picture and the phenomenological model discussed in the next section, our observation that this conditional dependence holds only for certain voltage sweep parameters may be consistent with one of the fluctuators becoming close to its zero-bias point where neither charge configuration is energetically preferred. Near the zero bias point, we expect that a TLF would exhibit enhanced sensitivity to bias perturbations.

\subsection{Phenomenological model}
\label{sec:phenomenological_model}
\noindent
Given the estimated transition rates, we would like to understand how the TLF energy scales vary with temperature, as well as whether there is an effect of heating due to the charge sensor as observed in previous work \cite{Ye202412}. Since the transition rates depend on the mechanism driving the TLF, in this section we consider the ratio of rates and assume detailed balance consistent with thermal equilibrium. Here, detailed balance corresponds to the condition $\Gamma_{10}/\Gamma_{01} = \exp{\left(-\Delta_{10}/k_{B} T\right)}$, where $\Gamma_{ij}$ is the attempt rate for transitioning from state $j$ to state $i$, $\Delta_{10} = E_{1}-E_{0}$ is the energy difference between TLF states $0$ and $1$, $T$ is the effective temperature of the bath to which the TLF is coupled, and $k_{B}$ is Boltzmann's constant.

As in our previous work \cite{Ye202412}, we consider the model for the TLF energy splitting and effective temperature due to charge sensor heating given by
\begin{eqnarray}
    \Delta_{10} & = & \overrightarrow{\lambda} \cdot (\mathbf{V} - \mathbf{V}_{\mathrm{Ref}}) \label{eqn:lever_arm} \\
    T & = & (T_{\mathrm{MC}}^{1+\beta} + \kappa Q)^{1/(1+\beta)}, \label{eqn:heating}
\end{eqnarray}
where in Eq. \ref{eqn:lever_arm} $\overrightarrow{\lambda}$ gives the ``lever arms'' between applied gate voltages (S,RP1) and the TLF energy splitting and $\mathbf{V}_{\mathrm{Ref}}$ is some reference voltage corresponding to the zero bias point for the TLF. In Eq. \ref{eqn:heating}, $T_{\mathrm{MC}}$ is the mixing chamber temperature of the dilution refrigerator, $Q$ is the charge sensor signal assumed to be proportional to the heating power shown in Fig. \ref{fig:CS_V_vs_RP1}(b), $\kappa$ is the heating ``lever arm'', and $\beta$ is the thermal conductivity exponent for the material. Given that the thermal conductivity exponent is poorly constrained, as in previous work we assume $\beta=3$ \cite{Ye202412}.

\begin{figure*}[ht!]
\centering
  \subfloat[Sweep of SET plunger gate RP1.]{\label{a}\includegraphics[width=85mm]{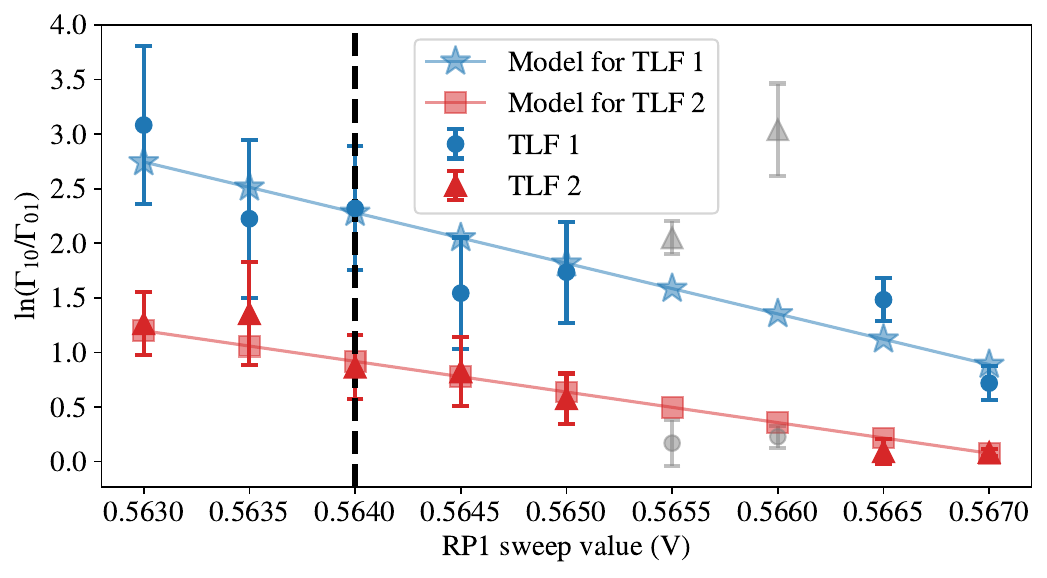}}\hfill
  \subfloat[Sweep of screening gate S.]{\label{b}\includegraphics[width=85mm]{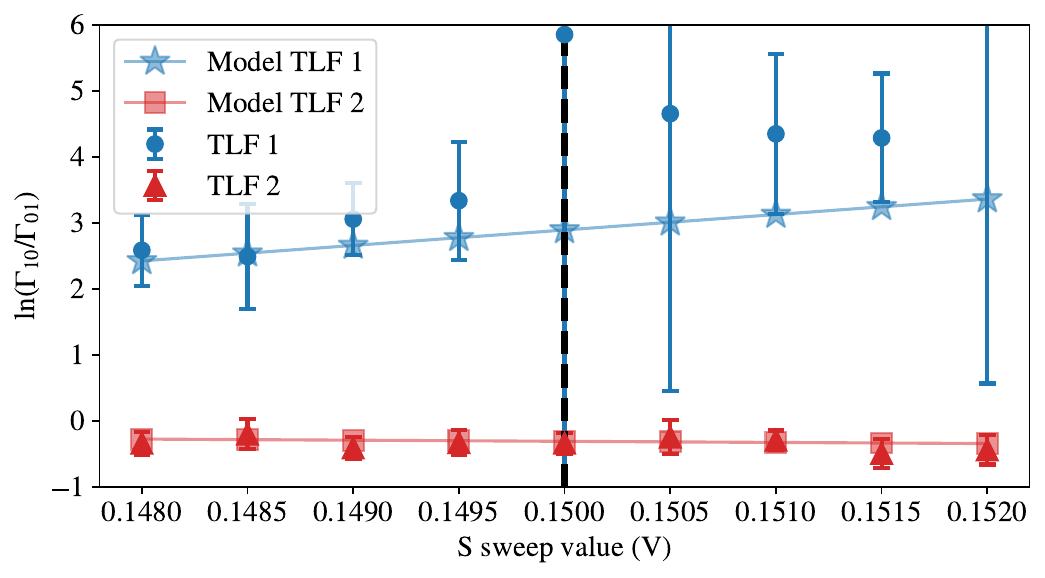}}\par
 \subfloat[Sweep of $T_{\mathrm{MC}}$.]{\label{c}\includegraphics[width=85mm]{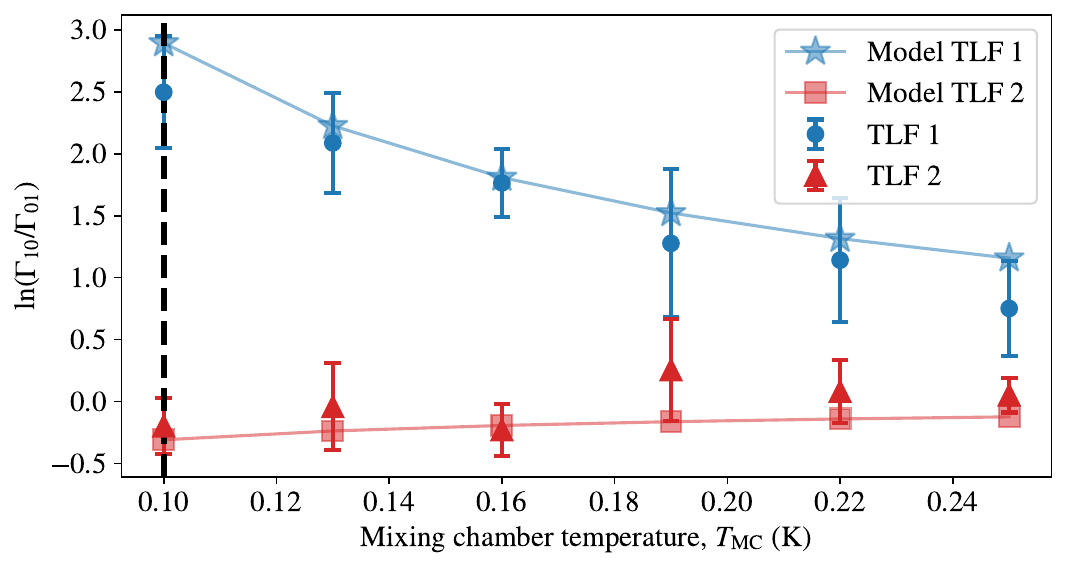}}
\caption{\textbf{Variation of the logarithm of the ratio of 2LF transition rates as a function of sweep parameter.} Values inferred by FHMM analysis of experimental data are given by the points with error bars, while the model fits are given by the curves. (a)
(b) (c). In all plots, the dashed vertical line corresponds to the fixed parameter value for which the other two sweeps were taken. In (a) the gray data points correspond to parameter values for which the transition
rates $\Gamma_{10}^{1,2}$ are comparable to or faster than the 60 Hz sample rate of data acquisition and likely suffer from significant aliasing error.
We exclude these points from the fits due to the poor estimation of the ratio of rates.}
\label{fig:ln_ratio_rates}
\end{figure*}

\begin{figure*}[ht!]
\centering
\begin{tabular}{cc}
  \includegraphics[width=75mm]{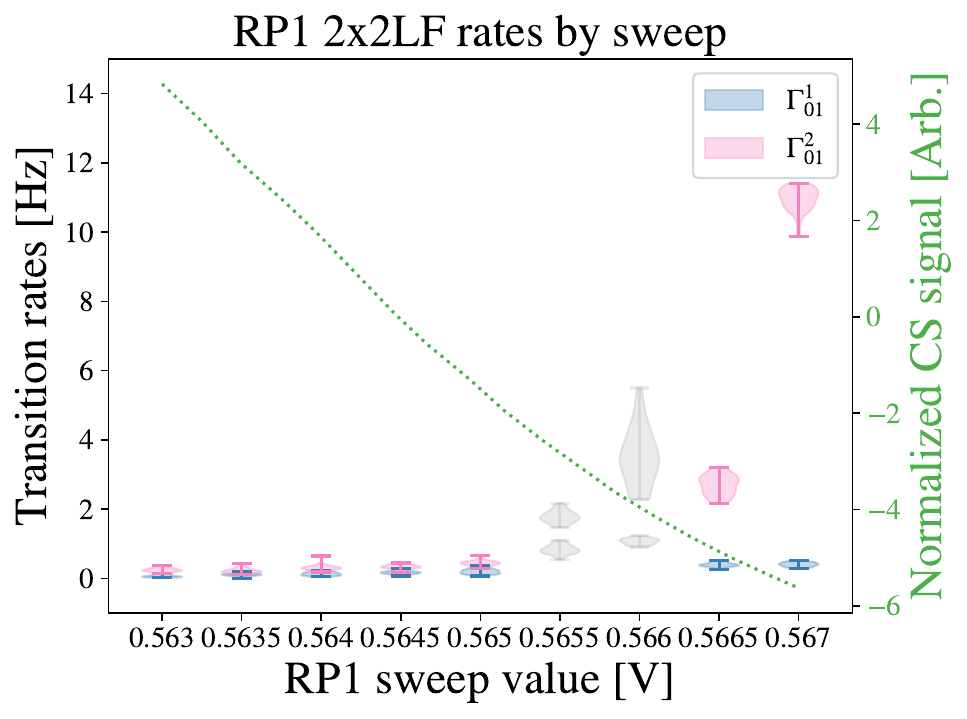} & 
  \includegraphics[width=75mm]{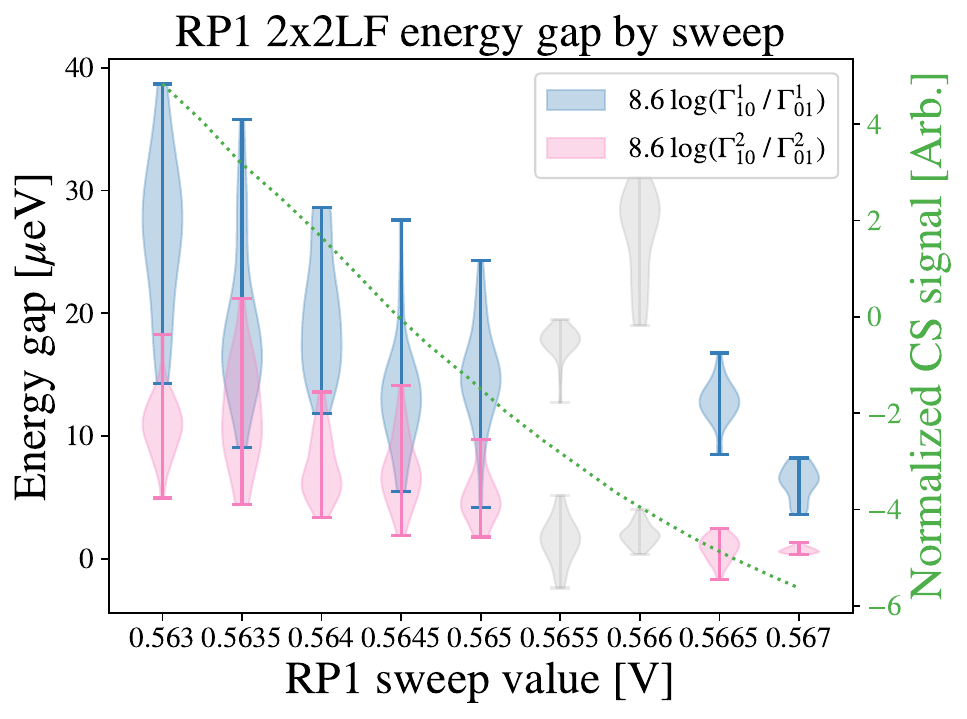}\\
  (a) & (b)  \\[6pt]
  \includegraphics[width=75mm]{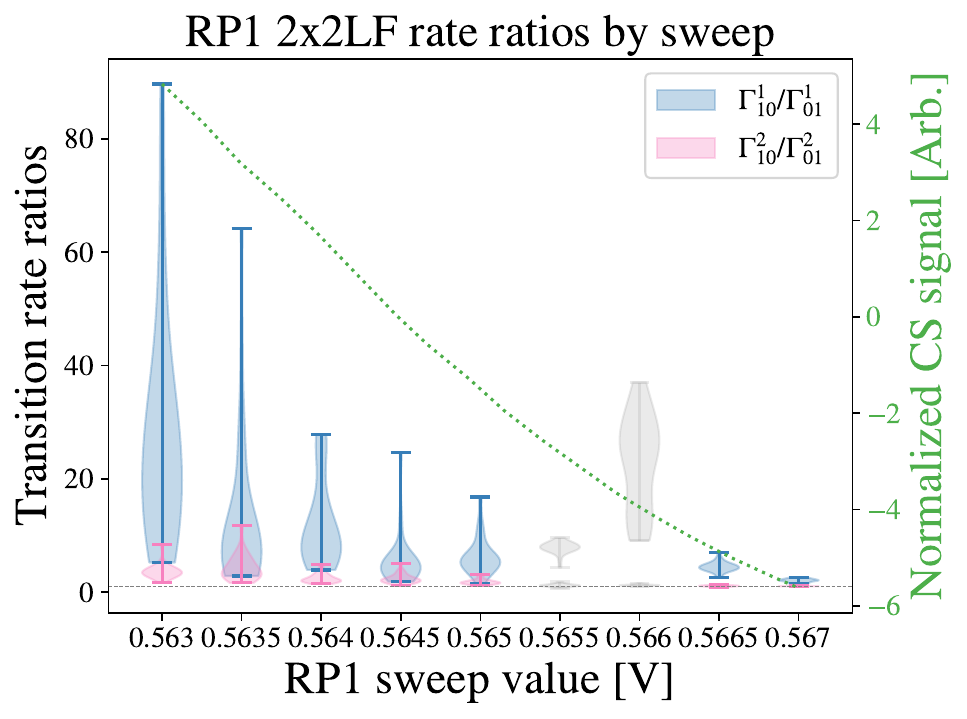} &
  \includegraphics[width=75mm]{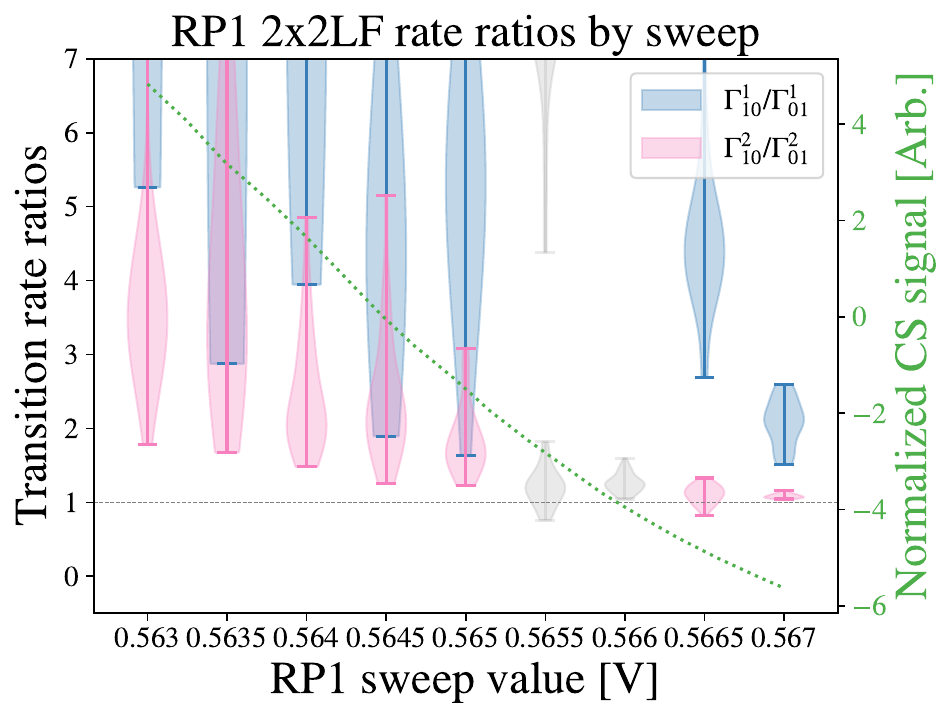} \\
  (c) Focusing on $\Gamma^{1}$  & (d) Focusing on $\Gamma^{2}$  \\[6pt]
\end{tabular}
\caption{\textbf{RP1 sweep FHMM rate fits for the 2x2LF model.}  The FHMM
fit rates for each sweep value, where we have grayed out scans where the
transition rates $\Gamma_{10}^{1,2}$ are comparable to or larger than the 60 Hz data acquisition sample rate and likely suffer from aliasing error.}
\label{fig:RP1_rates}
\end{figure*}

The fits of the above model to the results of our FHMM-based analysis of the experimental data are shown in Fig.~\ref{fig:ln_ratio_rates}, with our estimated lever arms given in Table \ref{tab:fit_params}. Here, we give further details on our fitting process. We minimize the misfit $M_{i} = \sum_{k} (r_{k,\mathrm{model}}^{i} - \mu_{k,\mathrm{data}}^{i})^{2}/(\sigma_{k,\mathrm{data}}^{i})^{2}$, where $\mu_{k,\mathrm{data}}^{i}$ and $\sigma_{k,\mathrm{data}}^{i}$ are the sample mean and standard deviation of $\ln(\Gamma_{10}^{j}/\Gamma_{01}^{j})$ as a function of parameter index $k$ for TLF $i$. 
Our reported uncertainties correspond to the parameter variation allowed within the 95\% confidence interval assuming the misfit $M_{i}$ to be $\chi^{2}$-distributed.
One feature of the data is that, while the S and $T_{\mathrm{MC}}$ sweeps are consistent relative to their common parameter value, the RP1 sweep is not. We suspect that this may be due to drift in the charge sensor calibration in the time between which the RP1 and S,$T_{\mathrm{MC}}$ datasets were taken, respectively. As a result, we treat the RP1 and (S,$T_{\mathrm{MC}}$) datasets separately. In addition, as shown in Fig. \ref{fig:RP1_rates}, our extracted transition rates $\Gamma_{10}^{1,2}$ for the RP1 sweep become similar to or larger than the measurement sample rate of 60 Hz for two of the parameter values. Since we expect this to lead to a significant aliasing error, we exclude these two points from our fit for the RP1 lever arms.

\begin{table}
    \begin{centering}
    \begin{tabular}{ccc}
    \toprule
    \textbf{Parameter} & \textbf{Name} & \textbf{Value ($\mathrm{ \mu eV/mV}$)} \\ \midrule
    Lever arm RP1 to TLF 1 & $\lambda_{RP1,1}$ &   $4.0 \pm 2.3$ \\
    Lever arm RP1 to TLF 2 & $\lambda_{RP1,2}$ & $2.4 \pm 1.3$\\
    Lever arm S to TLF 1 & $\lambda_{S,1}$ & -2.0$^*$ \\
    Lever arm S to TLF 2 & $\lambda_{S,2}$ & 0.15$^*$ \\
    \bottomrule
    \end{tabular}
    \end{centering}
   \caption{Estimated lever arms between RP1 and S gate electrodes to TLFs 1 and 2. For the lever arms to electrode RP1 we provide the 95\% confidence intervals, while for the lever arms to the screening gate S, $\lambda_{S,i}$, we provide only best-fit estimates since they are poorly constrained by the data (see main text and Fig. \ref{fig:uncertainty_S_TMC_sweeps}). }\label{tab:fit_params}
\end{table}

Our fits suggest that the heating lever arm $\kappa$ is small or negligible, since the best fit value for $\kappa$ in the RP1, S, and $T_{\mathrm{MC}}$ sweeps is zero. Hence, we may conclude that the majority of the variation in the ratio of transition rates is due to variation of the gate electrode voltages applied to RP1 or S. We find that the lever arms to the RP1 gate for both TLFs is on the order of a few $\mathrm{\mu eV/mV}$, as shown in the 95\% confidence region in Fig. \ref{fig:uncertainty_RP1_sweep}, which is consistent with our analysis of a TLF present in a different device \cite{Ye202412}. However, we find that the lever arms to the screening gate S are poorly constrained by the data, as shown in the 95\% confidence region of Fig. \ref{fig:uncertainty_S_TMC_sweeps}. The primary source of uncertainty in the lever arm to gate S arises from uncertainty in the location of the zero bias point of the TLF. Were the zero bias point to be identifiable, e.g. through performing a sweep over a larger range of S gate voltages, this would significantly constrain the lever arm.

\section{Conclusions}
In this work, we have observed multi-level charge noise in a Si/SiGe quantum dot device as a function of temperature and a variety of swept voltages. We analyzed these noise timeseries to differentiate contributions from background drift, a slow two-level fluctuator, and the remaining multi-level charge fluctuation. We modeled the remaining multi-level charge fluctuation time series using factorial hidden Markov modeling, to extract transition rates. We then fit these rates with a detailed balance model to extract individual charge fluctuator lever arms.

In order to mitigate the effects of drift on our Markov model, and to remove a nuisance fluctuator from the data, we implemented KLD and KDE-based drift detection and change point detection algorithms to assist in manually preprocessing the data. This allowed us to proceed with FHMM modeling of the remaining fluctuators in a targeted fashion.

We performed model comparison and validation on the remaining three prominent levels of the time series across the different sweeps, comparing 2x2LF, 3LF, and 4LF FHMM models, finding the 2x2LF model to be preferred writ large across the datasets. We fit a phenomenological, detailed balance model to the 2x2LF rate data yielding lever arms from TLF 1 to RP1 of $4.0\pm2.3\mu$eV/mV, from TLF 2 to RP1 of $2.4\pm1.3\mu$eV/mV, from TLF 1 to S of $-2.0\mu$eV/mV, and from TLF 2 to S of $0.15\mu$eV/mV.  (The latter two values are not well constrained.)

In the high RP1 voltage configuration, we found that the 4LF model outperformed the other models, and fit quite close to an embedded 2x2LF model. The detectable deviation from the embedded 2x2LF model revealed an emergent conditional rate dependence between the two nominally independent TLFs. We expect that one TLF would exhibit enhanced sensitivity to bias perturbations and thus could be more easily influenced by flipping of the second TLF, if the first TLF is near its zero bias point in this particular voltage configuration. This suggests that the two TLFs might be in close proximity to each other and may provide constraints on triangulating their locations.

\section*{Acknowledgements}
We thank Lisa F. Edge of HRL Laboratories for growing the heterostructure and Elliot J. Connors for fabricating the device used in this work.
This article has been co-authored by an employee of National Technology \& Engineering Solutions of Sandia, LLC under Contract No. DE-NA0003525 with the U.S. Department of Energy (DOE). The employee owns all right, title and interest in and to the article and is solely responsible for its contents. The United States Government retains and the publisher, by accepting the article for publication, acknowledges that the United States Government retains a non-exclusive, paid-up, irrevocable, world-wide license to publish or reproduce the published form of this article or allow others to do so, for United States Government purposes. The DOE will provide public access to these results of federally sponsored research in accordance with the DOE Public Access Plan \url{https://www.energy.gov/downloads/doe-public-access-plan.} This work was performed, in part, at the Center for Integrated Nanotechnologies, an Office of Science User Facility operated for the U.S. Department of Energy (DOE) Office of Science.

\bibliographystyle{unsrt}
\bibliography{multilevel}

@article{PhysRevApplied.13.024019,
  title = {Rapid High-Fidelity Spin-State Readout in $\mathrm{Si}$/$\mathrm{Si}$-$\mathrm{Ge}$ Quantum Dots via rf Reflectometry},
  author = {Connors, Elliot J. and Nelson, JJ and Nichol, John M.},
  journal = {Phys. Rev. Appl.},
  volume = {13},
  issue = {2},
  pages = {024019},
  numpages = {9},
  year = {2020},
  month = {Feb},
  publisher = {American Physical Society},
  doi = {10.1103/PhysRevApplied.13.024019}
}

@misc{Rojas-Arias2025,
	archiveprefix = {arXiv},
	author = {Juan S. Rojas-Arias and Akito Noiri and Jun Yoneda and Peter Stano and Takashi Nakajima and Kenta Takeda and Takashi Kobayashi and Giordano Scappucci and Seigo Tarucha and Daniel Loss},
	eprint = {2505.05875},
	primaryclass = {cond-mat.mes-hall},
	title = {Inferring charge noise source locations from correlations in spin qubits},
	url = {https://arxiv.org/abs/2505.05875},
	year = {2025}}

@article{Wang2025,
	author = {Wang, Zhanning and Gholizadeh, Sina and Hu, Xuedong and Das Sarma, S. and Culcer, Dimitrie},
	doi = {10.1103/PhysRevB.111.155403},
	issue = {15},
	journal = {Phys. Rev. B},
	month = {Apr},
	numpages = {12},
	pages = {155403},
	publisher = {American Physical Society},
	title = {Dephasing of planar Ge hole spin qubits due to $1/f$ charge noise},
	url = {https://link.aps.org/doi/10.1103/PhysRevB.111.155403},
	volume = {111},
	year = {2025}}

@article{Burkard2023,
	author = {Burkard, Guido and Ladd, Thaddeus D. and Pan, Andrew and Nichol, John M. and Petta, Jason R.},
	doi = {10.1103/RevModPhys.95.025003},
	issue = {2},
	journal = {Rev. Mod. Phys.},
	month = {Jun},
	numpages = {58},
	pages = {025003},
	publisher = {American Physical Society},
	title = {Semiconductor spin qubits},
	url = {https://link.aps.org/doi/10.1103/RevModPhys.95.025003},
	volume = {95},
	year = {2023}}

@article{Li2018,
	author = {Li, Zuo and Sotto, Mo{\"\i}se and Liu, Fayong and Husain, Muhammad Khaled and Yoshimoto, Hiroyuki and Sasago, Yoshitaka and Hisamoto, Digh and Tomita, Isao and Tsuchiya, Yoshishige and Saito, Shinichi},
	date = {2018/01/10},
	doi = {10.1038/s41598-017-18579-1},
	id = {Li2018},
	isbn = {2045-2322},
	journal = {Scientific Reports},
	number = {1},
	pages = {250},
	title = {Random telegraph noise from resonant tunnelling at low temperatures},
	url = {https://doi.org/10.1038/s41598-017-18579-1},
	volume = {8},
	year = {2018}}

@article{Shehata2023,
	author = {Shehata, M. Mohamed El Kordy and Simion, George and Li, Ruoyu and Mohiyaddin, Fahd A. and Wan, Danny and Mongillo, Massimo and Govoreanu, Bogdan and Radu, Iuliana and De Greve, Kristiaan and Van Dorpe, Pol},
	doi = {10.1103/PhysRevB.108.045305},
	issue = {4},
	journal = {Phys. Rev. B},
	month = {Jul},
	numpages = {28},
	pages = {045305},
	publisher = {American Physical Society},
	title = {Modeling semiconductor spin qubits and their charge noise environment for quantum gate fidelity estimation},
	url = {https://link.aps.org/doi/10.1103/PhysRevB.108.045305},
	volume = {108},
	year = {2023}}

@mastersthesis{Malcolm2020,
	author = {Malcolm, AJ},
	school = {University of Waterloo},
	title = {Multi-level Random Telegraph Noise Analysis Using Machine Learning Techniques},
	year = {2020}}

@article{Liu2018,
	author = {Liu, Fayong and Ibukuro, Kouta and Husain, Muhammad Khaled and Li, Zuo and Hillier, Joseph and Tomita, Isao and Tsuchiya, Yoshishige and Rutt, Harvey and Saito, Shinichi},
	journal = {Nanotechnology},
	number = {47},
	pages = {475201},
	publisher = {IOP Publishing},
	title = {Manipulation of random telegraph signals in a silicon nanowire transistor with a triple gate},
	volume = {29},
	year = {2018}}

@article{Machlup1954,
	author = {Machlup, Stefan},
	journal = {Journal of Applied Physics},
	number = {3},
	pages = {341--343},
	publisher = {American Institute of Physics},
	title = {Noise in semiconductors: spectrum of a two-parameter random signal},
	volume = {25},
	year = {1954}}

@article{Graaf2015,
	author = {de Graaf, S. E. and Danilov, A. V. and Kubatkin, S. E.},
	date = {2015/11/24},
	doi = {10.1038/srep17176},
	id = {de Graaf2015},
	isbn = {2045-2322},
	journal = {Scientific Reports},
	number = {1},
	pages = {17176},
	title = {Coherent interaction with two-level fluctuators using near field scanning microwave microscopy},
	url = {https://doi.org/10.1038/srep17176},
	volume = {5},
	year = {2015}}

@article{Yoneda2018,
	author = {Yoneda, Jun and Takeda, Kenta and Otsuka, Tomohiro and Nakajima, Takashi and Delbecq, Matthieu R. and Allison, Giles and Honda, Takumu and Kodera, Tetsuo and Oda, Shunri and Hoshi, Yusuke and Usami, Noritaka and Itoh, Kohei M. and Tarucha, Seigo},
	date = {2018/02/01},
	doi = {10.1038/s41565-017-0014-x},
	id = {Yoneda2018},
	isbn = {1748-3395},
	journal = {Nature Nanotechnology},
	number = {2},
	pages = {102--106},
	title = {A quantum-dot spin qubit with coherence limited by charge noise and fidelity higher than 99.9{\%}},
	url = {https://doi.org/10.1038/s41565-017-0014-x},
	volume = {13},
	year = {2018}}

@article{Wuetz2023,
	author = {Paquelet Wuetz, Brian and Degli Esposti, Davide and Zwerver, Anne-Marije J. and Amitonov, Sergey V. and Botifoll, Marc and Arbiol, Jordi and Sammak, Amir and Vandersypen, Lieven M. K. and Russ, Maximilian and Scappucci, Giordano},
	date = {2023/03/13},
	doi = {10.1038/s41467-023-36951-w},
	id = {Paquelet Wuetz2023},
	isbn = {2041-1723},
	journal = {Nature Communications},
	number = {1},
	pages = {1385},
	title = {Reducing charge noise in quantum dots by using thin silicon quantum wells},
	url = {https://doi.org/10.1038/s41467-023-36951-w},
	volume = {14},
	year = {2023}}

@article{Yoneda2023,
	author = {Yoneda, J. and Rojas-Arias, J. S. and Stano, P. and Takeda, K. and Noiri, A. and Nakajima, T. and Loss, D. and Tarucha, S.},
	date = {2023/12/01},
	doi = {10.1038/s41567-023-02238-6},
	id = {Yoneda2023},
	isbn = {1745-2481},
	journal = {Nature Physics},
	number = {12},
	pages = {1793--1798},
	title = {Noise-correlation spectrum for a pair of spin qubits in silicon},
	url = {https://doi.org/10.1038/s41567-023-02238-6},
	volume = {19},
	year = {2023}}

@article{Mehmandoost2024,
	author = {Mehmandoost, M. and Dobrovitski, V. V.},
	doi = {10.1103/PhysRevResearch.6.033175},
	issue = {3},
	journal = {Phys. Rev. Res.},
	month = {Aug},
	numpages = {18},
	pages = {033175},
	publisher = {American Physical Society},
	title = {Decoherence induced by a sparse bath of two-level fluctuators: Peculiar features of $1/f$ noise in high-quality qubits},
	url = {https://link.aps.org/doi/10.1103/PhysRevResearch.6.033175},
	volume = {6},
	year = {2024}}

@article{Rojas-Arias2023,
	author = {Rojas-Arias, J.S. and Noiri, A. and Stano, P. and Nakajima, T. and Yoneda, J. and Takeda, K. and Kobayashi, T. and Sammak, A. and Scappucci, G. and Loss, D. and Tarucha, S.},
	doi = {10.1103/PhysRevApplied.20.054024},
	issue = {5},
	journal = {Phys. Rev. Appl.},
	month = {Nov},
	numpages = {13},
	pages = {054024},
	publisher = {American Physical Society},
	title = {Spatial noise correlations beyond nearest neighbors in ${}^{28}\mathrm{Si}/$Si-Ge spin qubits},
	url = {https://link.aps.org/doi/10.1103/PhysRevApplied.20.054024},
	volume = {20},
	year = {2023}}

@article{Seedhouse2025,
	author = {Seedhouse, Amanda E. and Stuyck, Nard Dumoulin and Serrano, Santiago and Gilbert, Will and Huang, Jonathan Yue and Hudson, Fay E. and Itoh, Kohei M. and Laucht, Arne and Lim, Wee Han and Yang, Chih Hwan and Tanttu, Tuomo and Dzurak, Andrew S. and Saraiva, Andre},
	date = {2025/04/01},
	doi = {10.1038/s41598-024-79553-2},
	id = {Seedhouse2025},
	isbn = {2045-2322},
	journal = {Scientific Reports},
	number = {1},
	pages = {11065},
	title = {Wavelet correlation noise analysis for qubit operation variable time series},
	url = {https://doi.org/10.1038/s41598-024-79553-2},
	volume = {15},
	year = {2025}}

@article{Mickelsen2023,
	author = {Mickelsen, D. L. and Carruzzo, Herv\'e M. and Yu, Clare C.},
	doi = {10.1103/PhysRevB.108.195307},
	issue = {19},
	journal = {Phys. Rev. B},
	month = {Nov},
	numpages = {13},
	pages = {195307},
	publisher = {American Physical Society},
	title = {Interacting two-level systems as a source of $1/f$ charge noise in quantum dot qubits},
	url = {https://link.aps.org/doi/10.1103/PhysRevB.108.195307},
	volume = {108},
	year = {2023}}

@article{Cheng2025,
	author = {Cheng, Guoting and Guo, Jing},
	journal = {APL Quantum},
	number = {1},
	publisher = {AIP Publishing},
	title = {Modeling correlated-noise in silicon spin qubit device},
	volume = {2},
	year = {2025}}

@article{Freeman2016,
	author = {Freeman, Blake M and Schoenfield, Joshua S and Jiang, HongWen},
	journal = {Applied Physics Letters},
	number = {25},
	publisher = {AIP Publishing},
	title = {Comparison of low frequency charge noise in identically patterned Si/SiO2 and Si/SiGe quantum dots},
	volume = {108},
	year = {2016}}

@article{Grasser2012,
	author = {Tibor Grasser},
	doi = {https://doi.org/10.1016/j.microrel.2011.09.002},
	issn = {0026-2714},
	journal = {Microelectronics Reliability},
	note = {2011 Reliability of Compound Semiconductors (ROCS) Workshop},
	number = {1},
	pages = {39-70},
	title = {Stochastic charge trapping in oxides: From random telegraph noise to bias temperature instabilities},
	url = {https://www.sciencedirect.com/science/article/pii/S0026271411004203},
	volume = {52},
	year = {2012}}

@article{Choi2024,
	author = {Choi, Yujun and Coppersmith, S. N. and Joynt, Robert},
	doi = {10.1103/PhysRevA.110.052408},
	issue = {5},
	journal = {Phys. Rev. A},
	month = {Nov},
	numpages = {14},
	pages = {052408},
	publisher = {American Physical Society},
	title = {Using stochastic resonance of two-level systems to increase qubit coherence times},
	url = {https://link.aps.org/doi/10.1103/PhysRevA.110.052408},
	volume = {110},
	year = {2024}}

@article{Ye202407,
	author = {Ye, Feiyang and Ellaboudy, Ammar and Nichol, John M},
	journal = {arXiv preprint arXiv:2407.05439},
	title = {Stabilizing an individual charge fluctuator in a Si/SiGe quantum dot},
	year = {2024}}

@article{Cowie2024,
	author = {Megan Cowie and Procopios C. Constantinou and Neil J. Curson and Taylor J. Z. Stock and Peter Gr{\"u}tter},
	doi = {10.1073/pnas.2404456121},
	eprint = {https://www.pnas.org/doi/pdf/10.1073/pnas.2404456121},
	journal = {Proceedings of the National Academy of Sciences},
	number = {44},
	pages = {e2404456121},
	title = {Spatially resolved random telegraph fluctuations of a single trap at the Si/SiO$_2$ interface},
	url = {https://www.pnas.org/doi/abs/10.1073/pnas.2404456121},
	volume = {121},
	year = {2024}}

@article{Uren1988,
	author = {Uren, M. J. and Kirton, M. J. and Collins, S.},
	doi = {10.1103/PhysRevB.37.8346},
	issue = {14},
	journal = {Phys. Rev. B},
	month = {May},
	numpages = {0},
	pages = {8346--8350},
	publisher = {American Physical Society},
	title = {Anomalous telegraph noise in small-area silicon metal-oxide-semiconductor field-effect transistors},
	url = {https://link.aps.org/doi/10.1103/PhysRevB.37.8346},
	volume = {37},
	year = {1988}}

@misc{Albrecht2023,
	archiveprefix = {arXiv},
	author = {Dylan Albrecht and N. Tobias Jacobson},
	eprint = {2311.00084},
	primaryclass = {stat.CO},
	title = {NoMoPy: Noise Modeling in Python},
	url = {https://arxiv.org/abs/2311.00084},
	year = {2023}}

@article{Nielsen2021,
	author = {Nielsen, Erik and Rudinger, Kenneth and Proctor, Timothy and Young, Kevin and Blume-Kohout, Robin},
	doi = {10.1088/1367-2630/ac20b9},
	journal = {New Journal of Physics},
	month = {sep},
	number = {9},
	pages = {093020},
	publisher = {IOP Publishing},
	title = {Efficient flexible characterization of quantum processors with nested error models},
	url = {https://dx.doi.org/10.1088/1367-2630/ac20b9},
	volume = {23},
	year = {2021}}

@article{Kingman1962,
	author = {Kingman, J. F. C.},
	date = {1962/01/01},
	doi = {10.1007/BF00531768},
	id = {Kingman1962},
	isbn = {1432-2064},
	journal = {Zeitschrift f{\"u}r Wahrscheinlichkeitstheorie und Verwandte Gebiete},
	number = {1},
	pages = {14--24},
	title = {The imbedding problem for finite Markov chains},
	url = {https://doi.org/10.1007/BF00531768},
	volume = {1},
	year = {1962}}

@article{Johansen1974,
	author = {Johansen, S{\o}ren},
	journal = {Journal of the London Mathematical Society},
	number = {2},
	pages = {345--351},
	publisher = {Wiley Online Library},
	title = {Some results on the imbedding problem for finite Markov chains},
	volume = {2},
	year = {1974}}

@inproceedings{Johansen1973,
	address = {Dordrecht},
	author = {Johansen, S{\o}ren},
	booktitle = {Geometric Methods in System Theory},
	editor = {Mayne, D. Q. and Brockett, R. W.},
	isbn = {978-94-010-2675-8},
	pages = {227--236},
	publisher = {Springer Netherlands},
	title = {The Imbedding Problem for Finite Markov Chains},
	year = {1973}}

@article{Baake2020,
	author = {Michael Baake and Jeremy Sumner},
	doi = {https://doi.org/10.1016/j.laa.2020.02.016},
	issn = {0024-3795},
	journal = {Linear Algebra and its Applications},
	pages = {262-299},
	title = {Notes on Markov embedding},
	url = {https://www.sciencedirect.com/science/article/pii/S0024379520300823},
	volume = {594},
	year = {2020}}

@article{Kreinin2001,
	author = {Kreinin, Alexander and Sidelnikova, Marina},
	journal = {Algo Research Quarterly},
	number = {1/2},
	pages = {23--40},
	publisher = {Citeseer},
	title = {Regularization algorithms for transition matrices},
	volume = {4},
	year = {2001}}

@article{Casanellas2023,
	author = {Marta Casanellas and Jes{\'u}s Fern{\'a}ndez-S{\'a}nchez and Jordi Roca-Lacostena},
	doi = {10.5565/PUBLMAT6712308},
	journal = {Publicacions Matem{\`a}tiques},
	number = {1},
	pages = {411 -- 445},
	publisher = {Universitat Aut{\`o}noma de Barcelona, Departament de Matem{\`a}tiques},
	title = {{The embedding problem for markov matrices}},
	url = {https://doi.org/10.5565/PUBLMAT6712308},
	volume = {67},
	year = {2023}}

@article{Davies2010,
	author = {E. Davies},
	doi = {10.1214/EJP.v15-733},
	journal = {Electronic Journal of Probability},
	number = {none},
	pages = {1474 -- 1486},
	publisher = {Institute of Mathematical Statistics and Bernoulli Society},
	title = {{Embeddable Markov Matrices}},
	url = {https://doi.org/10.1214/EJP.v15-733},
	volume = {15},
	year = {2010}}

@article{Ye202412,
	author = {Ye, Feiyang and Ellaboudy, Ammar and Albrecht, Dylan and Vudatha, Rohith and Jacobson, N. Tobias and Nichol, John M.},
	doi = {10.1103/PhysRevB.110.235305},
	issue = {23},
	journal = {Phys. Rev. B},
	month = {Dec},
	numpages = {12},
	pages = {235305},
	publisher = {American Physical Society},
	title = {Characterization of individual charge fluctuators in Si/SiGe quantum dots},
	url = {https://link.aps.org/doi/10.1103/PhysRevB.110.235305},
	volume = {110},
	year = {2024}}

\begin{appendix}

\section{Charge sensor normalization}
\begin{figure*}[ht!]
\centering
\begin{tabular}{cc}
  \includegraphics[width=90mm]{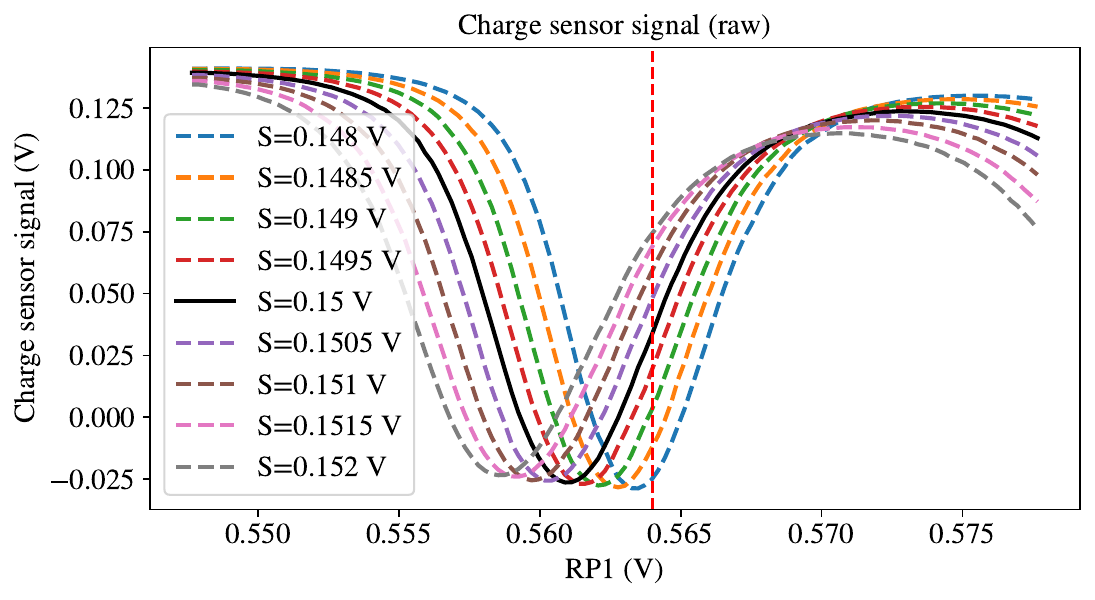} & 
  \includegraphics[width=90mm]{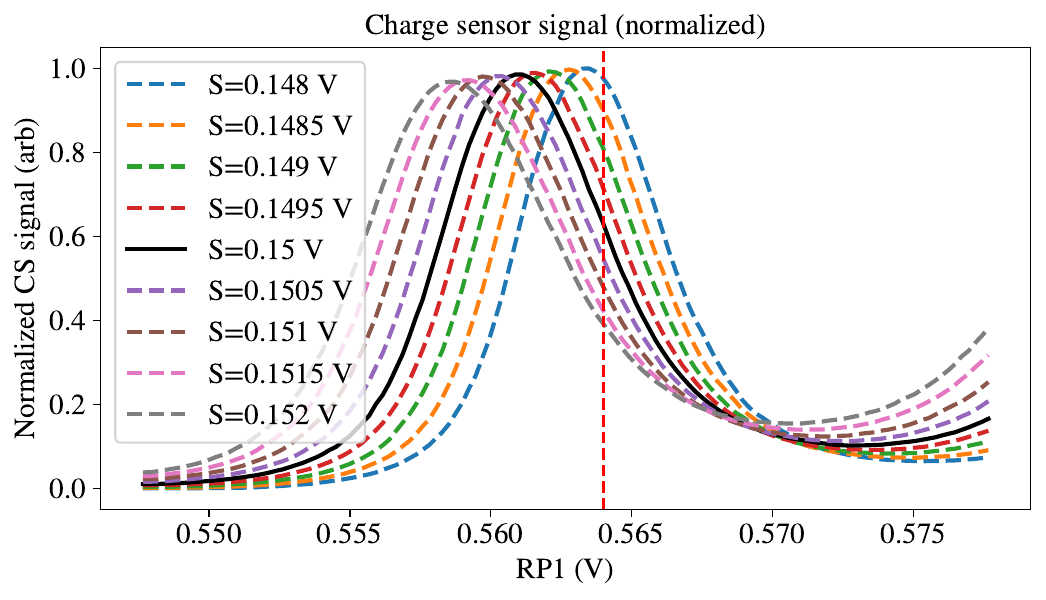}\\
  (a) & (b)  \\[6pt]
\end{tabular}
\caption{\textbf{Charge sensor} (a) Raw charge sensor signal as a function of SET plunger gate voltage RP1. (b) Normalized CS signal as used in the heating model. In both plots, the dashed vertical line corresponds to the fixed RP1 voltage for sweeps of other parameters such as S voltage and mixing chamber temperature $T_{\mathrm{MC}}$. Similarly, the solid black curve corresponds to the value of S held fixed when other parameters are swept.}
\label{fig:CS_V_vs_RP1}
\end{figure*}

The measured rf reflectometry signal for the charge sensor is shown in Fig. \ref{fig:CS_V_vs_RP1}(a) as a function of the two controlled gate voltages S and RP1. To map this to an effective charge sensor signal $Q$ that we treat in our heating model of Eq. \ref{eqn:heating} as proportional to heating through the charge sensor, we invert and normalize this signal as shown in Fig. \ref{fig:CS_V_vs_RP1}(b).

\begin{figure*}[ht!]
\centering
\begin{tabular}{cc}
  \includegraphics[width=90mm]{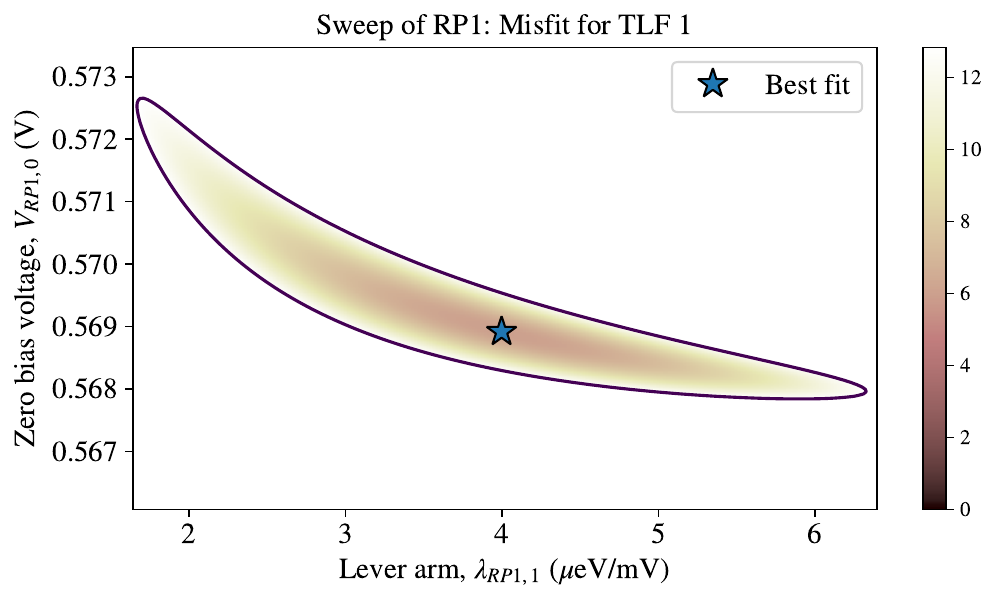} & 
  \includegraphics[width=90mm]{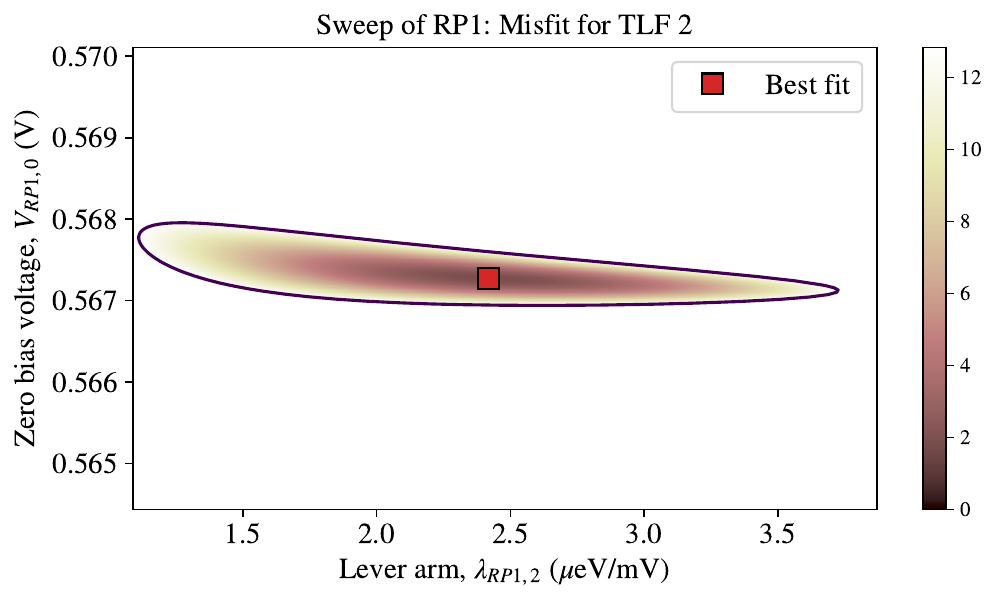}\\
  (a) & (b)  \\[6pt]
\end{tabular}
\caption{\textbf{Uncertainty in estimated zero-bias voltage point and lever arms for TLFs 1 and 2 for RP1 sweep.} Confidence region (95\% level) for (a) TLF 1 and (b) TLF 2.}
\label{fig:uncertainty_RP1_sweep}
\end{figure*}

\begin{figure*}[ht!]
\centering
\begin{tabular}{cc}
  \includegraphics[width=90mm]{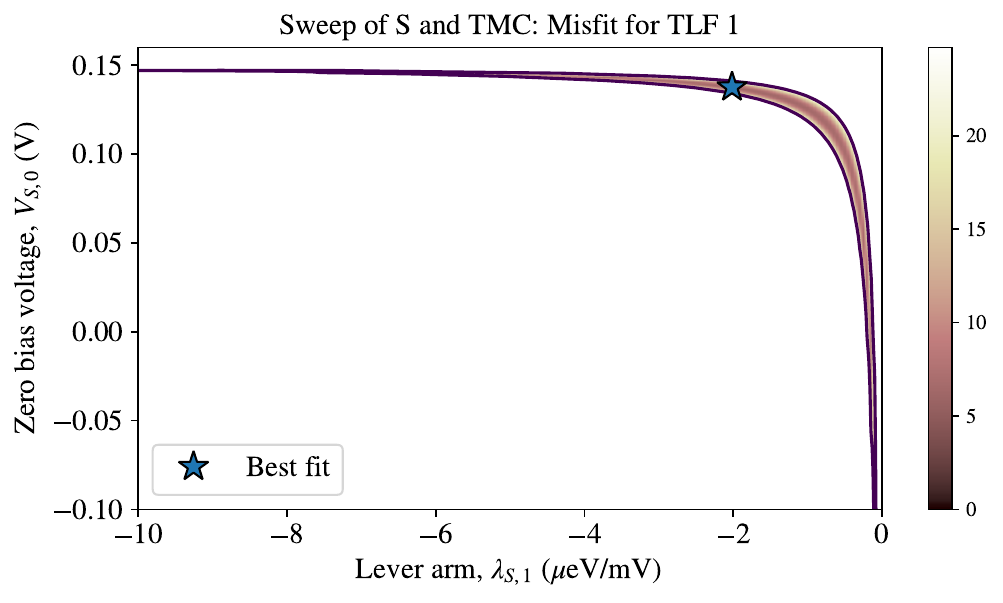} & 
  \includegraphics[width=90mm]{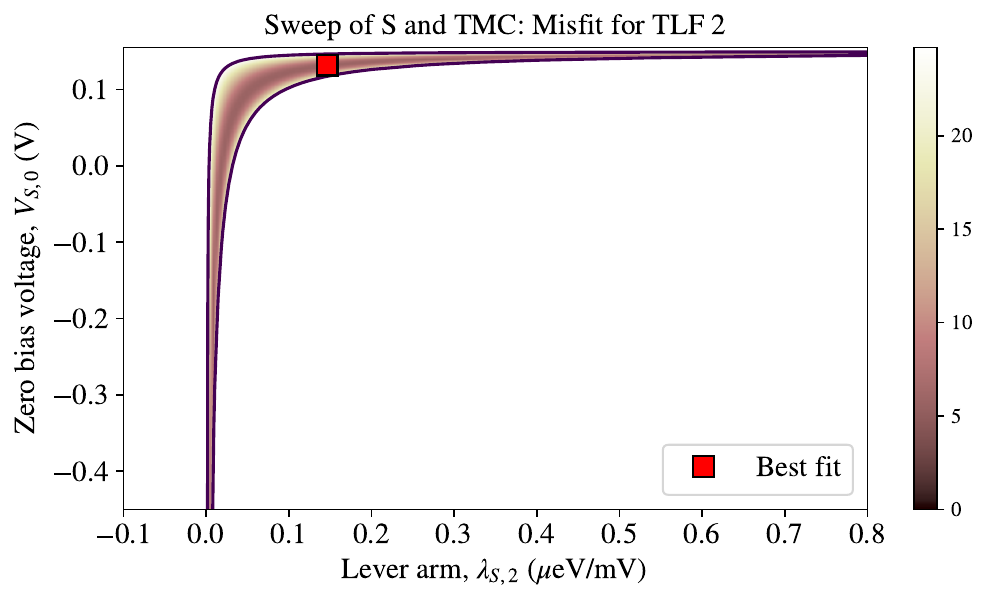}\\
  (a) & (b)  \\[6pt]
\end{tabular}
\caption{\textbf{Uncertainty in estimated zero-bias voltage point and lever arms for TLFs 1 and 2 for S and $T_{\mathrm{MC}}$ sweeps.} Confidence region (95\% level) for (a) TLF 1 and (b) TLF 2.}
\label{fig:uncertainty_S_TMC_sweeps}
\end{figure*}

\end{appendix}

\end{document}